\documentclass[letterpaper,useAMS,usenatbib]{mn2e}

\usepackage[utf8]{inputenc}
\usepackage{nicefrac}
\usepackage{latexsym}
\usepackage{amsmath, amssymb, mathrsfs, bm}
\usepackage{eqnarray}
\usepackage{url}
\usepackage{cases}
\usepackage{epsfig}
\usepackage{color}
\usepackage{graphicx}
\usepackage{grffile}
\usepackage{float}
\usepackage{soul}
\usepackage{enumerate}
\usepackage{multirow}
\usepackage{subcaption}

\usepackage{bm}
\usepackage{times}

\usepackage{etoolbox}
\preto\align{\par\nobreak\small\noindent}
\expandafter\preto\csname align*\endcsname{\par\nobreak\small\noindent}
\preto\multline{\par\nobreak\small\noindent}
\expandafter\preto\csname align*\endcsname{\par\nobreak\small\noindent}
\preto\flalign{\par\nobreak\small\noindent}
\expandafter\preto\csname align*\endcsname{\par\nobreak\small\noindent}
\preto\eqnarray{\par\nobreak\small\noindent}
\expandafter\preto\csname align*\endcsname{\par\nobreak\small\noindent}

\newcommand{\reffg}[1]{Fig.~(\ref{#1})}
\newcommand{\reftb}[1]{Tab.~(\ref{#1})}
\newcommand{\refeq}[1]{Eq.~(\ref{#1})}
\newcommand{\refsc}[1]{Sect.~(\ref{#1})}

\newcolumntype{L}[1]{>{\raggedright\arraybackslash}p{#1}}
\newcolumntype{C}[1]{>{\centering\arraybackslash}p{#1}}
\newcolumntype{R}[1]{>{\raggedleft\arraybackslash}p{#1}}

\def\hi{\textsc{Hi~}}

\def\dd{{\rm d}\,}
\def\sinc{{\rm sinc}}

\title[MeerKAT 1/f noise analysis]{\hi intensity mapping with MeerKAT: 1/f noise analysis}
\author[Yichao Li et. al.]{Yichao Li$^1$\thanks{dr.yichao.li@gmail.com}, 
Mario G. Santos$^{1,2}$, Keith Grainge$^3$, Stuart Harper$^3$, Jingying Wang$^1$\\
$^{1}$Department of Physics and Astronomy, University of the Western Cape, 
Robert Sobukwe Road, Belville 7535, South Africa\\
$^{2}$South African Radio Astronomy Observatory (SARAO), 2 Fir Street, Observatory, Cape Town, 7925, South Africa\\
$^{3}$ Jodrell Bank Centre for Astrophysics, Department of Physics and Astronomy,
The University of Manchester, Manchester M13 9PL, U.K\\
}

\begin{document}

\date{}

\pagerange{\pageref{firstpage}--\pageref{lastpage}} \pubyear{20??}

\maketitle

\label{firstpage}

\begin{abstract}
    The nature of the time correlated noise component (the 1/f noise) of single dish radio telescopes is critical to the detectability of  the HI signal in intensity mapping experiments.
    In this paper, we present the 1/f noise properties of the MeerKAT receiver system using South Celestial Pole (SCP) tracking data. We estimate both the temporal power spectrum density and the 2D power spectrum density for each of the antennas and polarizations.
    We apply Singular Value Decomposition (SVD) to the dataset and show that, by removing the strongest components, the 1/f noise can be drastically reduced, indicating that it is highly correlated in frequency.
    Without SVD mode subtraction, the knee frequency over a $20\,$MHz integration 
    is higher than $0.1\,\rm Hz$; with just $2$~mode subtraction, the knee frequency 
    is reduced to $\sim 3\times 10^{-3}\,{\rm Hz}$, 
    indicating that the system induced 1/f-type variations are 
    well under the  thermal noise fluctuations over a few hundred seconds time scales. 
    The 2D power spectrum shows that the 1/f-type variations are restricted to a small region in the time-frequency space, either with long wavelength correlations in frequency or in time. This gives a wide range of cosmological scales where the 21cm signal can be measured without further need to calibrate the gain time fluctuations. Finally, we demonstrate that a simple power spectrum parameterization is sufficient to describe the data and provide fitting parameters for both the 
    1D and 2D power spectrum.
\end{abstract}


\section{Introduction}

A major goal of modern cosmology is to understand the formation and
evolution of the cosmological large-scale structure (LSS), as well as
the information it carries from the early universe.
In the past decades, cosmologists have traced the LSS fluctuations
with wide-field spectroscopic and photometric surveys of galaxies
\citep{2005MNRAS.362..505C,2005ApJ...633..560E,2014MNRAS.441...24A,2017MNRAS.464.4807H}.
However, these surveys are often limited with either 
cosmologically small volumes or lower sampling density.
Furthermore, detecting individual objects at high significance is
time consuming.

Recently, the $21{\rm cm}$ emission line of Neutral Hydrogen (\textsc{Hi})
hyperfine spin-flip transition, has been proposed as another cosmological probe
of the LSS \citep[e.g.][]{2004MNRAS.355.1339B,2006ApJ...653..815M,2012RPPh...75h6901P}.
Instead of observing the \hi emission line from individual galaxies,
cosmologists proposed to measure the total \hi intensity of the galaxies within 
large voxels, a technique known as \hi intensity mapping (IM)
\citep{2008PhRvL.100i1303C,2008PhRvL.100p1301L,
2008PhRvD..78b3529M,2008PhRvD..78j3511P,2008MNRAS.383..606W,2008MNRAS.383.1195W,
2009astro2010S.234P,2010MNRAS.407..567B,2010ApJ...721..164S,2011ApJ...741...70L,
2012A&A...540A.129A,2013MNRAS.434.1239B}.
Because of the low angular resolution requirement, an \hi IM survey can be
quickly carried out with single dishes and extended to very large survey 
volumes. The \hi IM technique was 
explored with the Green Bank Telescope (GBT),
by measuring the cross-correlation function between an \hi IM survey and a
optical galaxy survey \citep{2010Natur.466..463C}. Later, the 
cross-correlation power spectrum between an \hi IM survey and an optical galaxy
survey was also reported with the GBT and Parkes telescopes
\citep{2013ApJ...763L..20M,2014atnf.prop.6273L,2018MNRAS.476.3382A,
2017MNRAS.464.4938W},
while the \hi IM auto power spectrum remains undetected \citep{2013MNRAS.434L..46S}.
There are several planned \hi IM experiments targeting the
post-reionization epoch, such as the Tianlai project \citep{2012IJMPS..12..256C}, the
Canadian Hydrogen Intensity Mapping Experiment (CHIME \cite{2014SPIE.9145E..22B}),
the Baryonic Acoustic Oscillations from Integrated Neutral Gas Observations 
(BINGO \cite{2013MNRAS.434.1239B}) and the
Hydrogen Intensity and Real-Time Analysis experiment (HIRAX \cite{2016SPIE.9906E..5XN}). 
The SKA has also been proposed as a major instrument to probe cosmology using 
this technique \citep{2015aska.confE..19S, 2015ApJ...803...21B, 2020PASA...37....7S}.
Recently, it was also proposed to have an \hi IM survey with the newly built
MeerKAT telescope in single-dish mode \citep{2017arXiv170906099S}.

There are several challenges for \hi IM power spectrum detection. 
The primary challenge is to remove the bright continuum radiation
of the Milky Way and extragalactic galaxies. The continuum radiation foreground
is known to have a smooth frequency spectrum and can be extracted by
fitting the spectrum with low order polynomial functions 
\citep{2012ApJ...744...29M}. However, due to instrumental effects,
the smooth-spectrum assumption breaks down and the foreground signal 
leaks into higher order fluctuation modes. Several foreground cleaning methods have been proposed to try to address this \citep{2014MNRAS.444.3183A,2015aska.confE..35W} 
and used in the analysis of GBT
and Parkes \hi IM survey \citep{2015ApJ...815...51S,2017MNRAS.464.4938W}.

\hi IM measurements also requires the receiver system to be stable.
However, the receiver system noise is known to have time correlated fluctuations, the
so-called 1/f-type noise (1/f noise). Such 1/f noise injects 
long-range correlations in time and leads to stripes in the final IM map. Since the measurements are performed in single dish mode (auto-correlation), they do not benefit from the suppression of 1/f noise afforded by interferometric measurements.
The 1/f noise effect has been discussed in previous analyses of 
Cosmic Microwave Background (CMB) experiments \citep{1996astro.ph..2009J}.
Several different destriping methods have been proposed and tested with
the analysis of CMB data  \citep{2002A&A...387..356M,2002A&A...391.1185S,
2004A&A...428..287K,2009A&A...506.1511K,2010MNRAS.407.1387S}.

The effect of 1/f noise on an \hi IM survey has been analysed through 
simulations \citep{2015MNRAS.454.3240B,2018MNRAS.478.2416H}. 
In the case of \hi IM, the data are collected 
across multiple frequency channels. However, the correlation of 1/f noise across
frequency is currently not very well understood. In this work, 
we develop a 1/f noise  power spectrum density estimator to extract 
the temporal and spectroscopic 1/f noise
properties of the MeerKAT receiver system using astronomical observation data.
We also apply Singular Value Decomposition (SVD) to the data
in order to reduce the 1/f-type fluctuations.
The paper is organized as followed. 
Our power spectrum density analysis method and the 1/f noise model are introduced
in \refsc{sec:model}; the details of observation data are given in
\refsc{sec:data}; the SVD method is introduced in \refsc{sec:svd};
a mask filling method is introduced in \refsc{sc:fm} to reconstruct the 
missing data due to the RFI flagging; the results are discussed in
\refsc{sec:results}; and the conclusions are summarized in \refsc{sec:conclusion}.

\section{1/f Noise Power Spectrum Density Model}\label{sec:model}
\subsection{Temporal Power Spectrum Density}
The time-ordered data (in arbitrary units) as a function of time $t$ and 
frequency $\nu$, $d(t,\nu$), can be modeled as the input temperature,
$T_{\rm in}(t, \nu)$, multiplied by the gain, $G(t, \nu)$:
\begin{align}\label{eqobs}
   d(t, \nu) = G(t, \nu) T_{\rm in}(t, \nu) + n(t,\nu),
\end{align}
where $n(t,\nu)$ represents the white noise term (a Gaussian variable uncorrelated in time and frequency).
The input temperature can be expressed as
$T_{\rm in} = \left( T_{\rm sky} + T_{\rm rx} \right)$,
where $T_{\rm sky}$ is the sky temperature (convolved by the telescope primary beam) 
and $T_{\rm rx}$ is the receiver temperature.
The gain, $G(t, \nu)$, refers to the gain of the amplifiers in the receiver. There are 
several sources of temporal fluctuations in \refeq{eqobs}. First, there are the usual 
sky fluctuations. Since in these observations the telescope is fixed and
pointing at the South Celestial Pole (SCP), we should not see much variation, 
except as a result of point sources or our galaxy
moving in and out of asymmetries in the primary beam (or rising and setting). 
These changes are expected to vary slowly over time. 
Second, we have instrumental fluctuations. There are slow gain drifts that we
expect to be able to calibrate out,
the intrinsic white noise fluctuations mentioned above that should average down in time and the
``non-calibrated'', correlated gain fluctuations that are the focus of this paper 
(the 1/f noise). Note
that even the long time scales gain drifts could in principle be incorporated 
in this correlated noise term although this is unnecessary as it can be calibrated
out. The correlated noise has simple statistical properties (at least on 
timescales $\lesssim 1$ h) which we describe next.

We start by defining, $G(t, \nu) \equiv \bar{G}(\nu) + \delta G(t, \nu)$ and 
$T_{\rm in}(t, \nu) \equiv \bar{T}_{\rm in}(\nu) + \delta T_{\rm in}(t, \nu)$, 
where $\bar{G}(\nu)$ and $\bar{T}_{\rm in}(\nu)$ are the time averaged quantities. 
We subtract and divide the data by its time average, $\bar{d}(\nu)$, 
taking only the varying part for the rest of the analysis: 
$\delta_d(t,\nu)\equiv\frac{d(t, \nu)}{\bar{d}(\nu)} - 1$. 
By dividing by $\bar{d}(\nu)$ we also cancel out the frequency dependence both from the sky and instrument (the part that is stable in time, e.g., the bandpass). To first order we can then write
\begin{align}\label{eq:dt}
    \delta_d(t,\nu) \approx \frac{\delta T_{\rm in}(t, \nu)}{\bar{T}_{\rm in}(\nu)}
             + \frac{\delta G(t, \nu)}{\bar{G}(\nu)} + \frac{n(t,\nu)}{\bar{T}_{\rm in}(\nu) \bar{G}(\nu)}.
\end{align}
The first term corresponds to sky fluctuations, which should be small, while 
$\delta G / G$ incorporates the 1/f-type fluctuations. 

We are going to assume that we can model the "non-calibrated" instrument fluctuations through a 
correlated Gaussian distribution (both in time and frequency).
These can be characterised by the temporal power spectrum density function,
$\hat{S}^{t}(f, \nu)$, which is estimated via the Fourier transfer of $\delta_d(t,\nu)$ along the time axis as
\begin{align}\label{eq:tcorr}
    \hat{S}^{t}(f, \nu) = \left| \sqrt{\frac{\delta t}{N_t}} \sum_{p=0}^{N_t-1} \delta_d(p \delta t, \nu)
    \exp[-2\pi i f p \delta t] \right|^2, 
\end{align}
in which, $f$ is the temporal frequency;
$\delta t$ is the time resolution of the data ($t=p\delta t$) and 
$N_t$ the number of time samples. 

If we only have the white noise fluctuations, we can write,
\begin{align}
    & S^t(f,\nu)\equiv \left<\hat{S}^{t}(f, \nu) \right> =
    \frac{\delta t}{N_t} \sum_{p=0}^{N_t-1} <\delta_d^2(p \delta t, \nu)> \\
    & = \delta t \frac{\sigma_n^2}{\bar{T}_{\rm in}^2(\nu) \bar{G}^2(\nu)}
    = \frac{\delta t}{\bar{T}_{\rm in}^2(\nu)}\frac{\bar{T}_{\rm in}^2(\nu)}{\delta t \delta\nu} = \frac{1}{\delta \nu},
\end{align}
where $\sigma_n^2$ 
is the white noise variance and $\sigma_n^2/\bar{G}^2(\nu) = \frac{T^2_{\rm in}}{\delta \nu \delta t}$ for a frequency
bin width of $\delta \nu$ and integration time of $\delta t$ \citep[e.g.][]{2009tra..book.....W}.
Please note that, since $\delta_d$ is normalized with the system temperature, the power spectrum 
density is normalized with $T_{\rm sys}^2$.
The power spectrum in the time direction from the extra 1/f-type noise 
component can then be modelled as \citep{2018MNRAS.478.2416H},
\begin{align}\label{eq:tcorrfn}
    S^t_{\rm fn}(f,\nu) = \frac{1}{\delta \nu}
    \left(\frac{f_k}{f}\right)^\alpha,
\end{align}
where $\alpha$ is the spectral index of the noise.
This enforces that for large $f$ the 1/f noise power spectrum goes to zero and 
the overall power spectrum becomes dominated by white noise.
Our model for the full temporal power spectrum density function is then,
\begin{align}\label{eq:tmodel}
    S^{t}(f,\nu) = \frac{A}{\delta \nu} \left( 1 + \left(\frac{f_k}{f}\right)^\alpha \right),
\end{align}
where $A$ ($\sim 1$) is as a free parameter fit together with
$\alpha$ and $f_k$.

\subsection{2-Dimensional Power Spectrum Density}

\begin{figure*}
    \small
    \centering
    \begin{minipage}[t]{0.48\textwidth}
    \includegraphics[width=\textwidth]{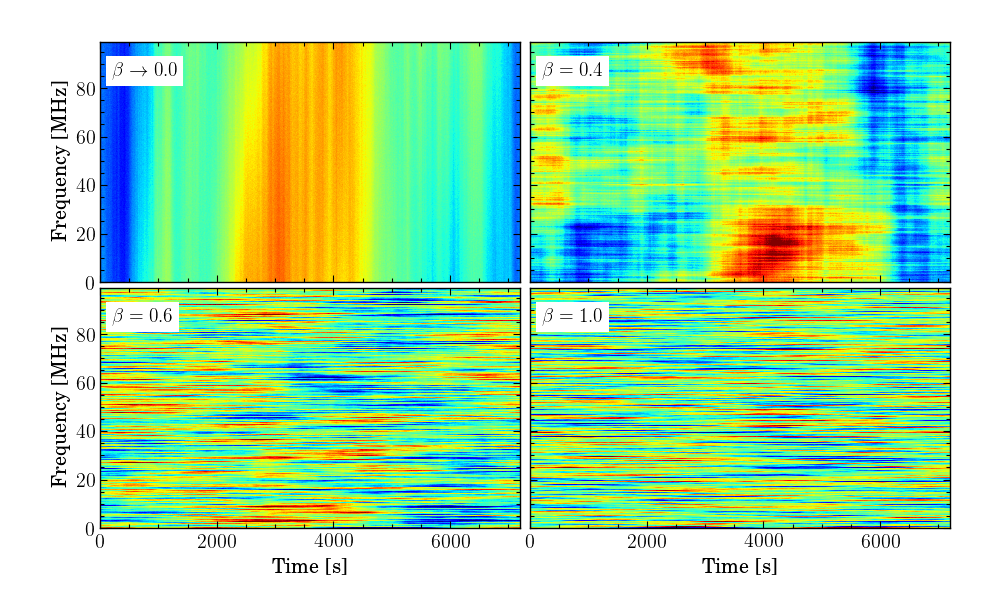}
    \subcaption{}\label{fig:sim_fn_wf}
    \end{minipage}
    \hfill
    \begin{minipage}[t]{0.48\textwidth}
    \includegraphics[width=\textwidth]{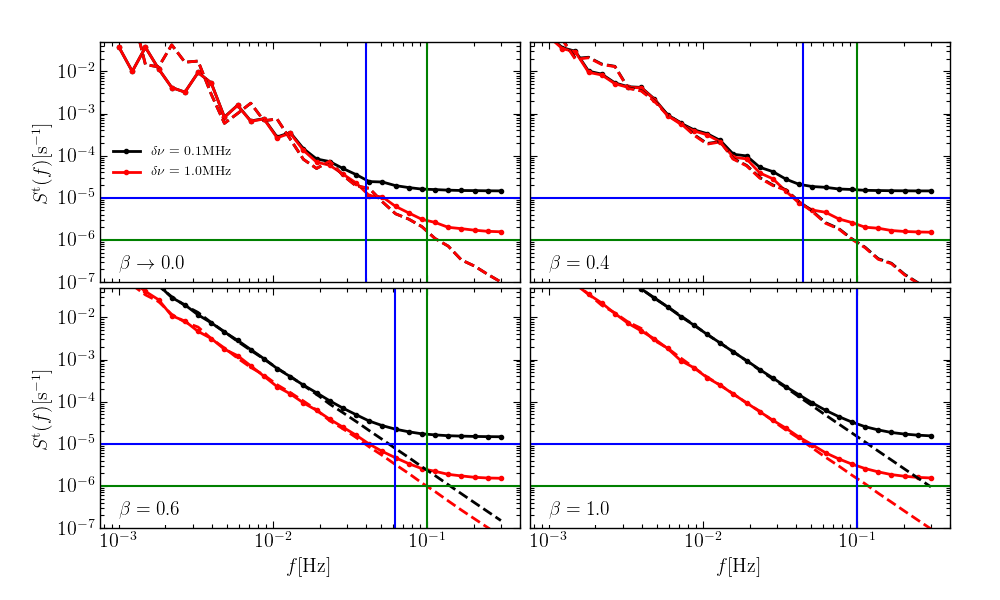}
    \subcaption{}\label{fig:sim_fn_ps}
    \end{minipage}
    \caption{
        {\bf (a)}:
        The waterfall plot of the simulated time-ordered data
        using the model of \refeq{eq:fn2d} as the input power spectrum.
        The simulation uses 1/f noise parameters: $f_k=0.1\,{\rm Hz}$ 
        at frequency resolution $\delta \nu=1\,{\rm MHz}$,
        $\alpha=2.5$ and different values of $\beta$ as shown in each panel. 
        $\beta=1$ corresponds to uncorrelated 1/f noise across frequency while 
        $\beta\to0$ is fully correlated (one can still see a fluctuation since we 
        cannot simulate $\beta=0$ exactly).
        {\bf (b)}:
        The temporal power spectrum density of the simulated 
        time-ordered data as shown in \reffg{fig:sim_fn_wf}. 
        The black dashed curve shows the power spectrum of pure 1/f noise simulation
        (set white noise equals to $0$) with frequency resolution of $0.1\,{\rm MHz}$
        and the solid curve shows the simulation with white noise added; 
        The red dashed/solid curve show the same simulations with frequency
        resolution reduced to $1\,{\rm MHz}$. 
        The horizontal lines indicate the white noise level with frequency
        resolution of $0.1\,{\rm MHz}$ (blue) and $1\,{\rm MHz}$ (green), respectively.
        The cross points with the vertical lines indicate the 
        knee frequency at the corresponding frequency resolution, which is estimated
        with \refeq{eq:fk2f0}.
        For $\beta=1$ (fully uncorrelated), we expect the 1/f noise power 
        spectrum to be inversely proportional to the frequency resolution.
    }
\end{figure*}

The 1/f noise can potentially be correlated in frequency. We need therefore 
to consider a 2-dimensional power spectrum to fully describe its statistics.
This 2-D power spectrum density can be estimated by Fourier transforming the
observed time-ordered data along the time and frequency axes, 
\begin{align}
\hat{S}(f, \tau) = \left| \sqrt{\frac{\delta t \delta \nu}{N_t N_\nu}}
\sum_{p=0}^{N_t-1} \sum_{k=0}^{N_\nu-1} \delta_d \exp \left[ -2\pi i 
\left( f p \delta t + \tau k \delta \nu \right) \right]
\right|^2,
\end{align}
in which $f$ is the temporal frequency and $\tau$ is the spectroscopic frequency 
(the Fourier conjugate in the frequency domain). In this case, if we only have white noise fluctuations, $\hat{S}(f, \tau)=1$.
We then build our 2-Dimensional (2D)
power spectrum density model as
\begin{align}\label{eq:2dtmp}
S_{\rm fn}(f, \tau) = F(f) H(\tau),
\end{align}
where $F(f)$ describes the temporal correlation power spectrum,
\begin{align}
F(f) = \frac{1}{\delta \nu}\left( \frac{f_k}{f} \right)^\alpha,
\end{align}
with $f_k$ the knee frequency defined at the frequency resolution of $\delta\nu$).
$H(\tau)$ is the spectroscopic correlation power spectrum density, which can be modeled
as \citep{2018MNRAS.478.2416H},
\begin{align}
H(\tau) = \left(\frac{\tau_0}{\tau}\right)^{\frac{1-\beta}{\beta}},
\end{align}
where $\beta$ specifies the amount of correlation across frequencies and $\tau_0=1/(N_{\nu}\delta\nu)$. 
Combining the white noise term, the 2D power spectrum model can be
expressed as,
\begin{align}\label{eq:fn2d}
S(f, \tau) = A \left( 1 + \frac{1}{K\delta \nu}\left(\frac{f_k}{f}\right)^\alpha 
 \left(\frac{\tau_0}{\tau}\right)^{\frac{1-\beta}{\beta}}\right),
\end{align}
in which, $K=\int \dd \tau \sinc^2(\pi\delta\nu\tau)
\left(\frac{\tau_0}{\tau}\right)^{(1-\beta)/\beta}$.
The derivation of \refeq{eq:fn2d} is shown in the appendix.
$A\sim 1$ due to the normalization with $T_{\rm sys}^2$.
In our analysis, $A$ is set as an overall amplitude parameter 
which can be constrained by the
observation data together with $f_0$, $\alpha$ and $\tau_0$.

The knee frequency $f_k$ as a function of frequency resolution is an important 
consideration 
for the LSS correlation signal on the largest scale sizes. 
For example, if we are
interested in line-of-sight scales of $\sim100\,{\rm Mpc}/h$, 
at $900\,{\rm MHz}$ (i.e. $z \sim 0.6$), this corresponds to frequency scales of $\sim25\,{\rm MHz}$. 
Depending upon the knee frequency, at the $\sim25\,{\rm MHz}$
frequency resolution 
there is the potential to detect
the 1/f noise more significantly
than at lower values of the frequency resolution.
The knee frequency at two different frequency resolutions, $\delta \nu$, $\delta \nu'$, is related via,
\begin{align}\label{eq:fk2f0}
\lg f_k = \lg f_{k'} + \frac{1}{\alpha}\lg 
\left( \frac{K\delta \nu}{K'\delta \nu'} \right).
\end{align}
The derivation is shown in the appendix. We test the shift of the knee frequency
with simulated time-ordered data.
\reffg{fig:sim_fn_wf} shows the waterfall plots of the simulated time-ordered
data with different frequency correlation properties.
As $\beta\to 0$, the 1/f noise becomes fully correlated over the frequency band. 
As $\beta\to 1$, the frequency correlation length is reduced
and the 1/f noise between different frequencies becomes independent 
(down to the frequency resolution).  

The corresponding temporal power spectrum of the simulated data
is shown in \reffg{fig:sim_fn_ps}. 
The black solid curve shows the power spectrum with $0.1\,{\rm MHz}$ 
frequency resolution, which is the raw frequency resolution of the simulation;
while the red curve shows the power spectrum after averaging over 
$10$ frequency channels. 
The dashed curves show the simulation with only 1/f noise 
(set white noise level to 0).
The horizontal lines indicate the white noise level with frequency
resolution of $0.1\,{\rm MHz}$ (blue) and $1\,{\rm MHz}$ (green), respectively.
The cross points with the vertical lines indicate the 
knee frequency at the corresponding frequency resolution, which is estimated
with \refeq{eq:fk2f0}.
The white noise floor, as expected, is 
reduced by one order of magnitude after frequency averaging. However, the 1/f noise
level behaves differently with different $\beta$ values. In the case
of $\beta=0$, the 1/f noise is fully correlated over the frequency band.
The level of 1/f noise power spectrum does not change with averaging 
frequency channels, but the white noise does. The different behavior
between 1/f noise and white noise results in a higher knee frequency value
at lower frequency resolution. With $\beta$ increasing, the 1/f noise
behaves more like the white noise. In the case of $\beta=1$, the 1/f noise
if fully uncorrelated between frequencies and the power spectrum 
level is reduced by one order of magnitude as well. In this case, the 
knee frequency does not change with frequency resolution.

\subsection{Parameter fitting}
The parameters that characterise the 1/f noise can be constrained by fitting the model against the
measured
noise power spectrum.
We build the $\chi^2$ function both for temporal and the 2-D power spectrum
density function,
\begin{align}
    \chi^2_{t} = \frac{\left(\langle\hat{S}^{t}(f)\rangle_\nu - S^{t}(f)\right)^2}
    {\sigma^2_{\hat{S}^{t}}}\;\; {\rm and},  \;\; 
    \chi^2 = \frac{\left(\hat{S}(f, \tau) - S(f, \tau)\right)^2}
    {\sigma^2_{\hat{S}}},
\end{align}
in which, $\langle\ \rangle_\nu$ represents the average over the frequency channels and
$\sigma_{\hat{S}^{t}}$, $\sigma_{\hat{S}}$ are 
the estimated errors of the temporal and 2-D power spectrum 
density, respectively.
The errors of the temporal power spectrum density
are estimated via the standard deviation of the power spectrum 
density using different frequency channels,
\begin{align}
    \sigma^2_{\hat{S}^{t}} = \frac{1}{N_\nu} \left(
    \langle ( \hat{S}^{t}(f) )^2 \rangle_\nu
    - ( \langle \hat{S}^{t}(f) \rangle_\nu )^2 \right),
\end{align}
where $N_\nu$ is the number of frequency channels.
The errors of the 2-D power spectrum density is simply estimated via,
\begin{align}
    \sigma^2_{\hat{S}} = \frac{1}{N_{\rm mode}} \hat{S}^2(f, \tau),
\end{align}
where $N_{\rm mode}$ is the number of Fourier modes within the $f-\tau$ bins.
We constrain the free parameters by minimizing the $\chi^2$ function,
using the publicly available Markov Chain Monte Carlo (MCMC) algorithm \texttt{emcee}
\citep{2019JOSS....4.1864F}.

\section{Observations and Data Reduction}\label{sec:data}

Details on the MeerKAT telescope can be found in \citet{2016mks..confE...1J}, \citet{2018ApJ...856..180C} and \citet{2020ApJ...888...61M}.
In order to characterise the 1/f-type fluctuations of the system noise, 
we need a constant input signal and a long duration observation.
Both requirements can be satisfied by tracking the 
SCP over several hours. Two SCP datasets are used 
in our analysis. One was collected in 2016 with a few antennas; 
the other was collected in 2019 using the majority of the MeerKAT array.

\begin{description}
\item {\bf 2016 SCP Data}
    The data in 2016 (SCP16) was observed with three of the MeerKAT antennas,
    named M017, M021 and M036, pointing at the South Celestial Pole (SCP).
    The observation started with a $20\,{\rm Hz}$ sampling
    rate for $3.5\,{\rm min}$ (Experiment ID 20160922-0004),
    followed by $1\,{\rm Hz}$ sampling rate for
    $2\,{\rm hours}$ (Experiment ID 20160922-0005).
        The frequencies range from $856\,{\rm MHz}$ to $1711.791\,{\rm MHz}$, 
    with $4096$ frequency channels and
    $0.209\,{\rm MHz}$ frequency resolution.
    
\item {\bf 2019 SCP Data}
    The SCP tracking data in 2019 (SCP19) was observed on April 24 
    (Experiment ID 20190424-0024)
    using $\sim 60$ antennas over $2.5\,{\rm hours}$.
    The data was taken with a $0.5\,{\rm Hz}$ sampling rate.
    The frequency range and resolution for the data in 2019 are the same as in 2016.
\end{description}

\begin{figure}
    \small
    \centering
    \includegraphics[width=0.48\textwidth]{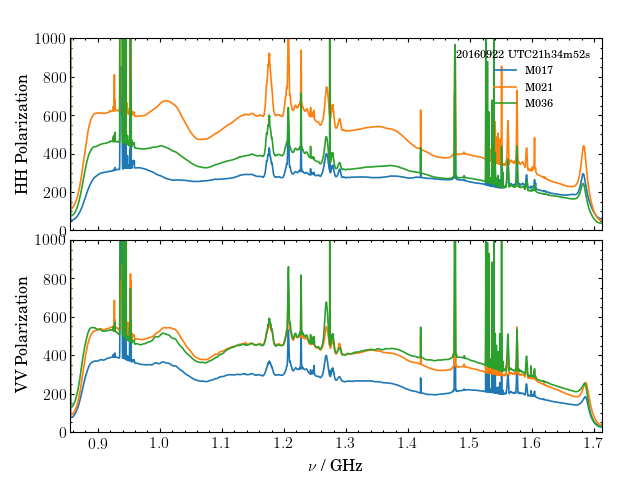}\hspace{-0.1cm}
    \includegraphics[width=0.48\textwidth]{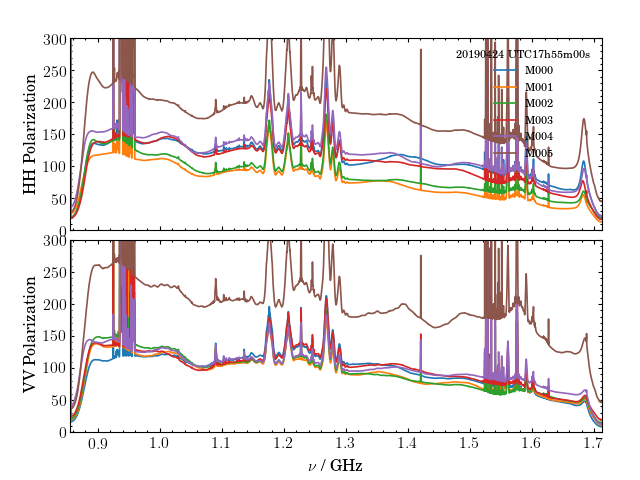}
    \caption{
        The frequency spectrum of SCP tracking data. The top panel shows the spectrum
        of the SCP16 20Hz data; the bottom panel shows the SCP19 data. 
        In each panel, the top/bottom sub-panels show the two polarizations.
        The y-axis is the uncalibrated raw detector output power.
    }\label{fig:spec}
\end{figure}

\reffg{fig:spec} shows the frequency spectrum of the SCP tracking data,
averaged over the observation time.
The top panel shows the spectrum of all three antennas used in the SCP16.
The bottom panel shows the spectrum of the first 6 antennas used in the
SCP19 observation. The amplitude shown here is the uncalibrated raw
detector output power. The scattering of the amplitude across antennas is mainly 
due to the different digital gain settings.
We use the relatively RFI-free frequency range between $1294.8672\,{\rm MHz}$ and
$1503.8516\,{\rm MHz}$ for the rest of the analysis
but exclude the HI signal from our galaxy at 1420MHz.

\begin{figure*}
    \small
    \centering
    \includegraphics[width=0.48\textwidth]{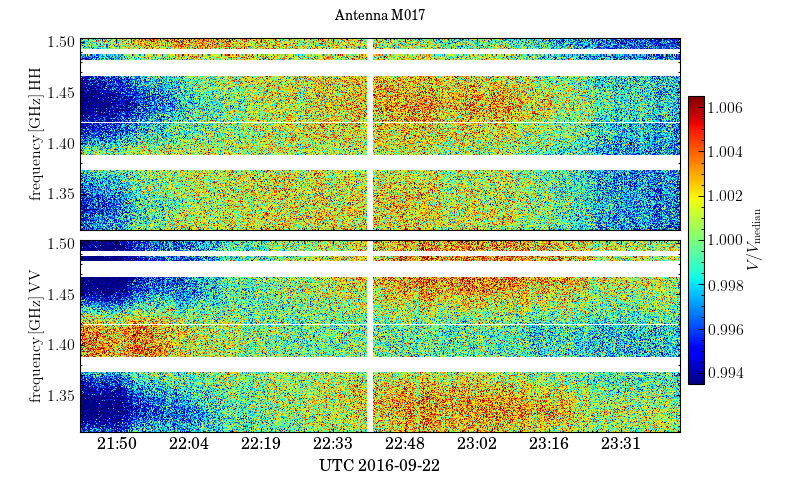}\hspace{-0.4cm}
    \includegraphics[width=0.48\textwidth]{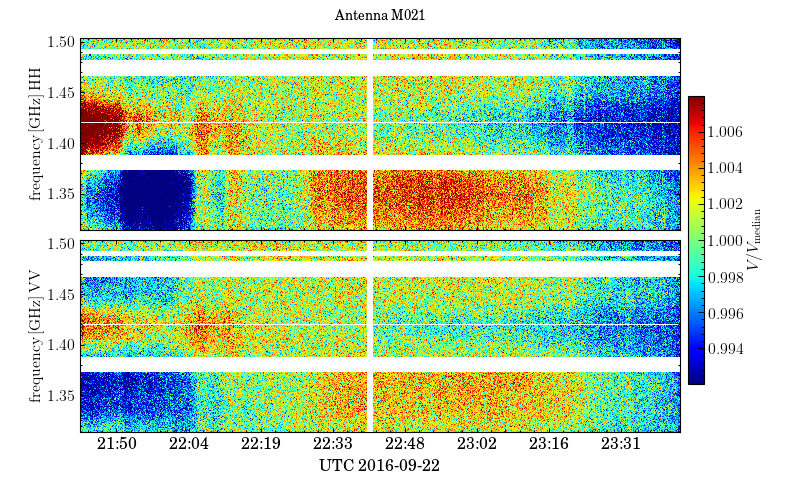}\hspace{-0.4cm}
    \includegraphics[width=0.48\textwidth]{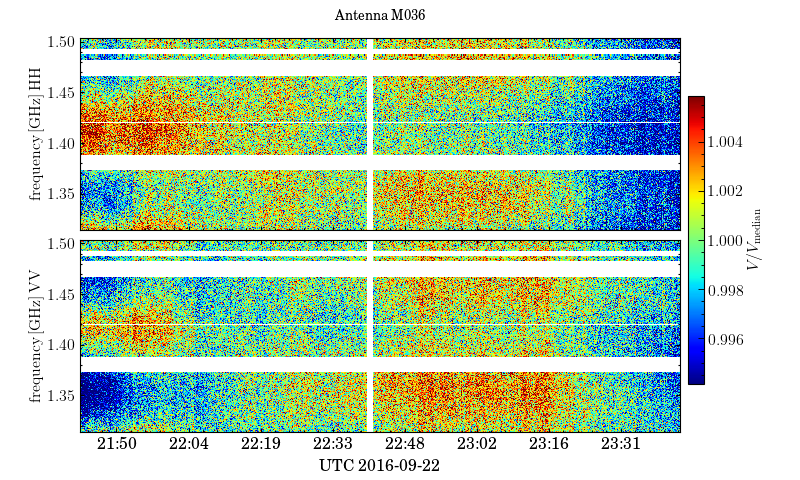}\hspace{-0.4cm}
    \includegraphics[width=0.48\textwidth]{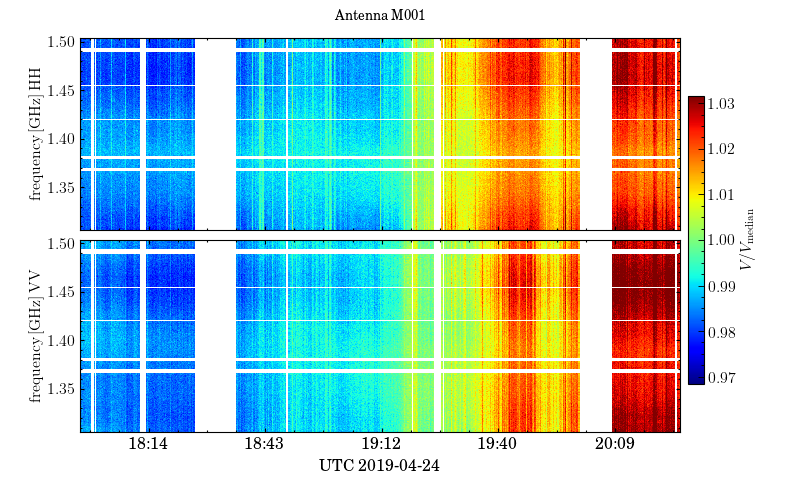}
    \caption{
        The waterfall plots of the SCP tracking data. The SCP16 $1\,{\rm Hz}$ 
        data observed by 
        antenna M017, M021 and M036 are shown in the top-left, top-right and
        bottom-left panels, respectively; The SCP19 data observed by M001, 
        as an example, is shown in the bottom-right panel ($0.5\,{\rm Hz}$ sampling rate). 
        The frequency range is truncated between $1294.8672\,{\rm MHz}$ and
        $1503.8516\,{\rm MHz}$, which is relatively RFI-free.
        The power amplitude of each frequency is normalized with the median values
        along the time axis.
        The color scale is restricted between the mean value of the
        shown data plus/minus $2$ times of the r.m.s. of the data.
    }\label{fig:wf}
\end{figure*}

As discussed in section \ref{sec:model}, we are interested in the fluctuations around the time average, $\delta_d(t,\nu) = \frac{d(t, \nu)}{\bar{d}(\nu)} - 1$. We then need to normalise the data by its time average. Note however that we use median values instead of the mean values to avoid time varying RFI. We expect the receiver temperature as well as most external sources to be constant during the observation time since the telescope is fixed and observing the SCP. Therefore, this normalisation should calibrate out the telescope bandpass as well as most spectral features from the sky, ground pick up and atmosphere. The remaining time and frequency fluctuations in $\delta_d(t,\nu)$ are expected to be from 1/f and white noise. However, some fluctuations can still be present from sources rising and setting and due to primary beam asymmetries.
The waterfall plots of the normalized data are shown in the \reffg{fig:wf}.

Three antennas from SCP16 at $1$Hz sampling rate 
are shown in the top-left, top-right and bottom-left panels of \reffg{fig:wf}, while 
one antenna from SCP19, as an example, is shown in the bottom-left panel. 
In each panel, the two polarizations
are shown in the top and bottom sub-panels. The color scale is restricted to run between
the mean value plus/minus $2$ times the r.m.s. of the data shown.
The amplitude varies over both frequency and observation time.
The variation of SCP16 data is around $0.3\%$ of the mean, $T_{\rm sys}$, 
which is much less than the SCP19 data. The SCP19 data shows strong variations in time while showing  
strong correlations across frequency. 
Such frequency-correlated variations are synchronous across
different antennas, which indicates an 
larger overall environment variation during the 2019 observation.

\section{Time-ordered SVD}\label{sec:svd}

\begin{figure*}
    \small
    \centering
    \includegraphics[width=0.45\textwidth]{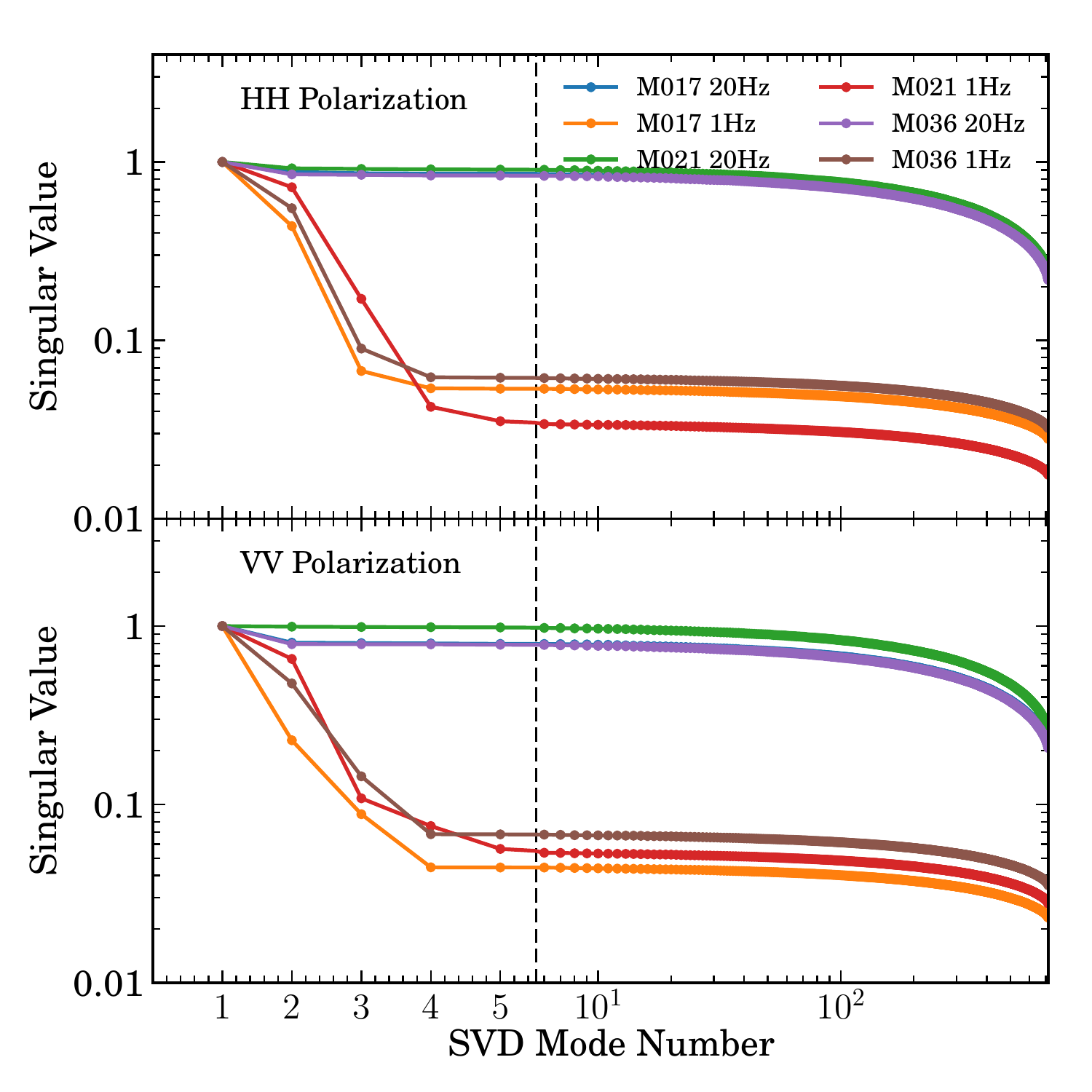}\hspace{-0.1cm}
    \includegraphics[width=0.45\textwidth]{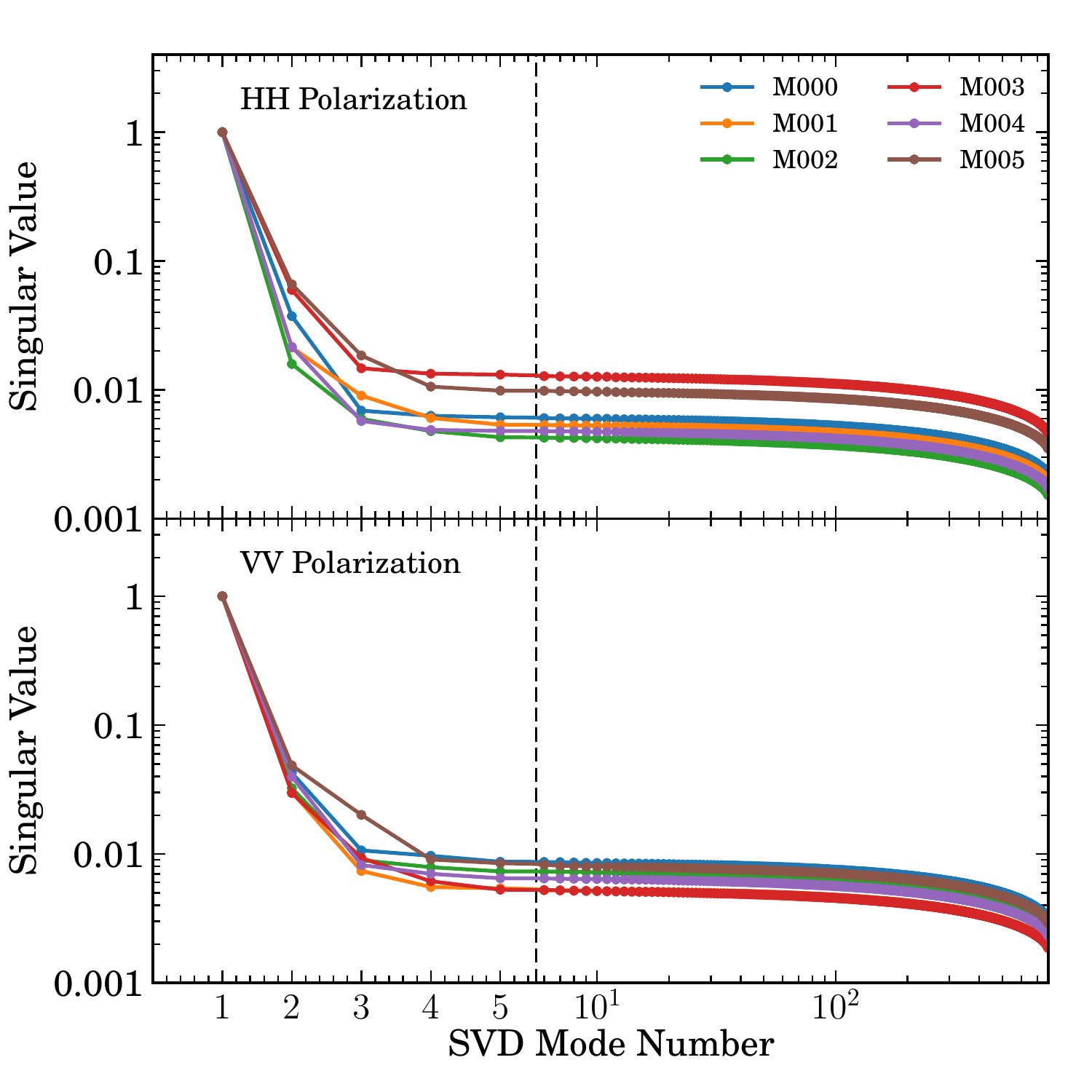}
    \caption{
        The singular values of each of the SVD modes, normalized to the
        first singular value. The black dashed line indicates the first 5 singular values.
        The left panel shows the singular values of the SCP16 data,
        both for $1\,{\rm Hz}$ and $20\,{\rm Hz}$ sampling rate,
        observed with different antennas.
        The right panels show the singular values of data from Apr. 24th,
        SCP19. The results of the first $6$ antennas
        are shown in different colors. For the $20\,{\rm Hz}$ data, the changes in amplitude are small since the modes are dominated by noise.
    }\label{fig:svdvalue}
\end{figure*}

\begin{figure*}
    \small
    \centering
    \includegraphics[width=0.45\textwidth]{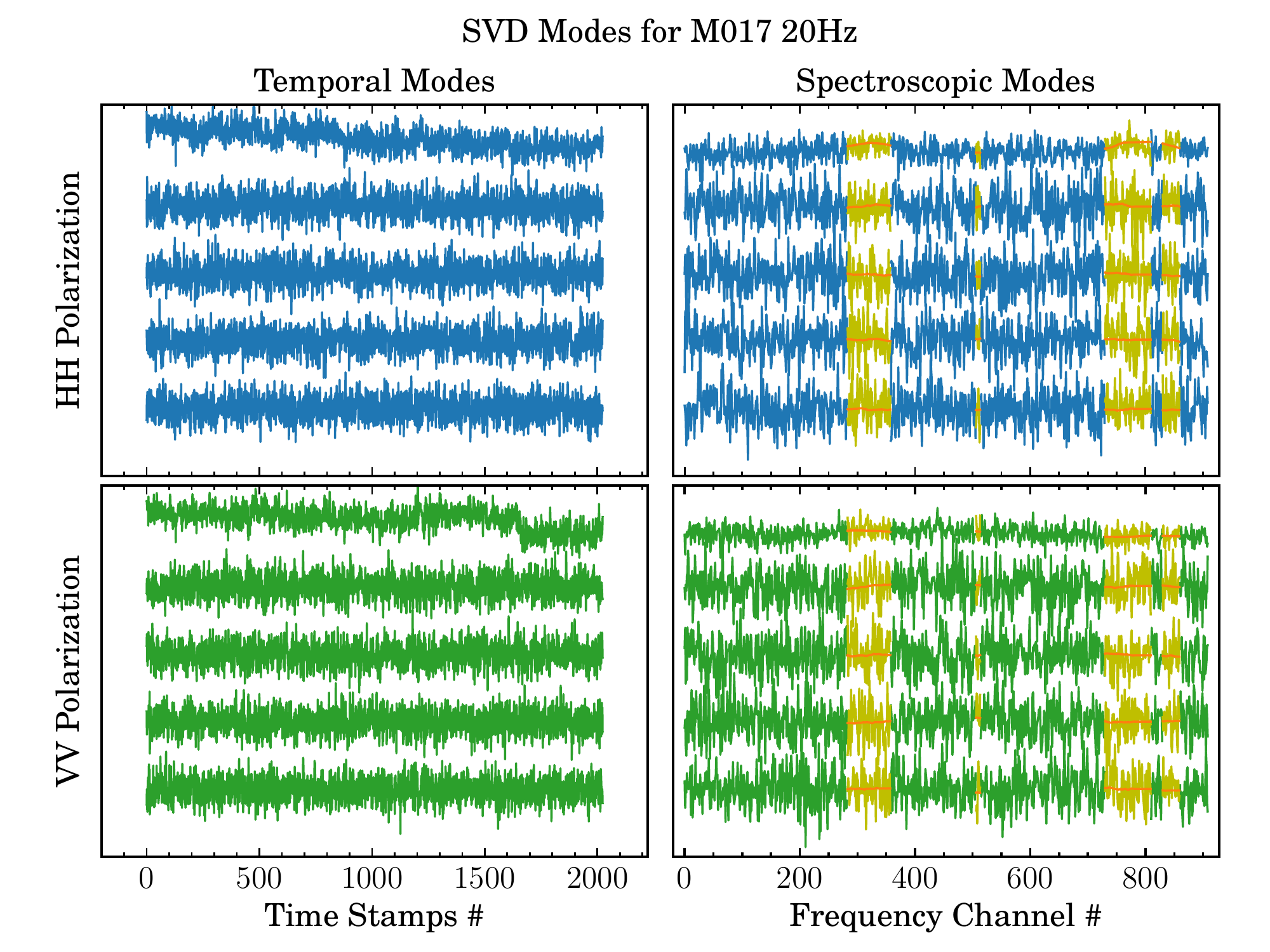}
    \includegraphics[width=0.45\textwidth]{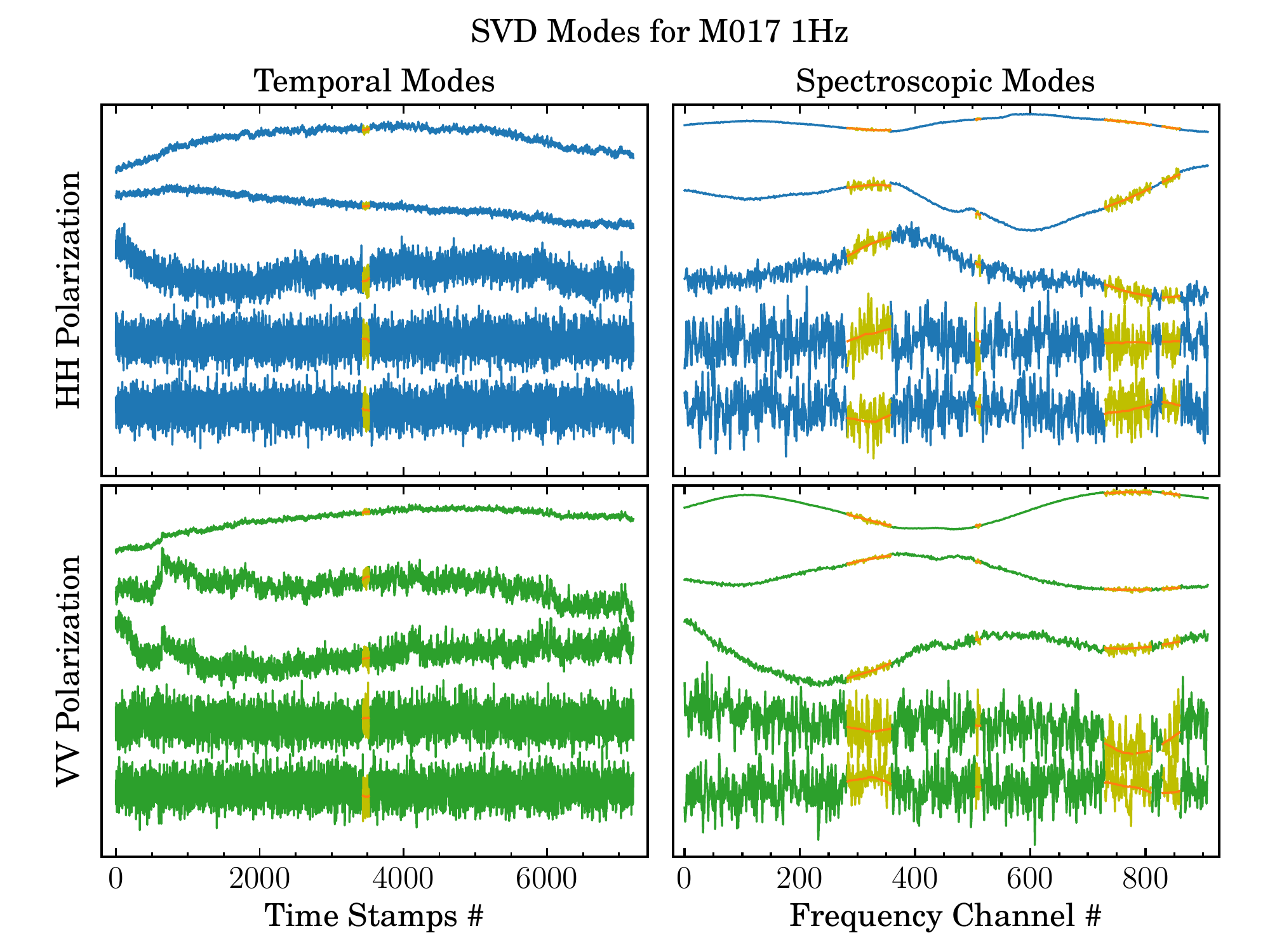}
    \caption{
        The singular modes of each of the first $5$ SVD modes, of antenna M017,
        as an example, of the SCP16 data, 
        with results of $20\,{\rm Hz}$ data on the left panel and
        $1\,{\rm Hz}$ data on the right.
        The HH and VV polarization are shown in the top and bottom sub-panels,
        respectively;
        and the temporal and spectroscopic modes are shown in the
        left and right sub-panels, respectively.
        The areas shown in yellow are the re-filled values due to the RFI flagging,
        and the orange curves are the Wiener filtered smooth curves.
        We discuss our mask-filling strategy in \refsc{sc:fm}.
    }\label{fig:svdmodes}
\end{figure*}

\begin{figure*}
    \small
    \centering
    \includegraphics[width=0.45\textwidth]{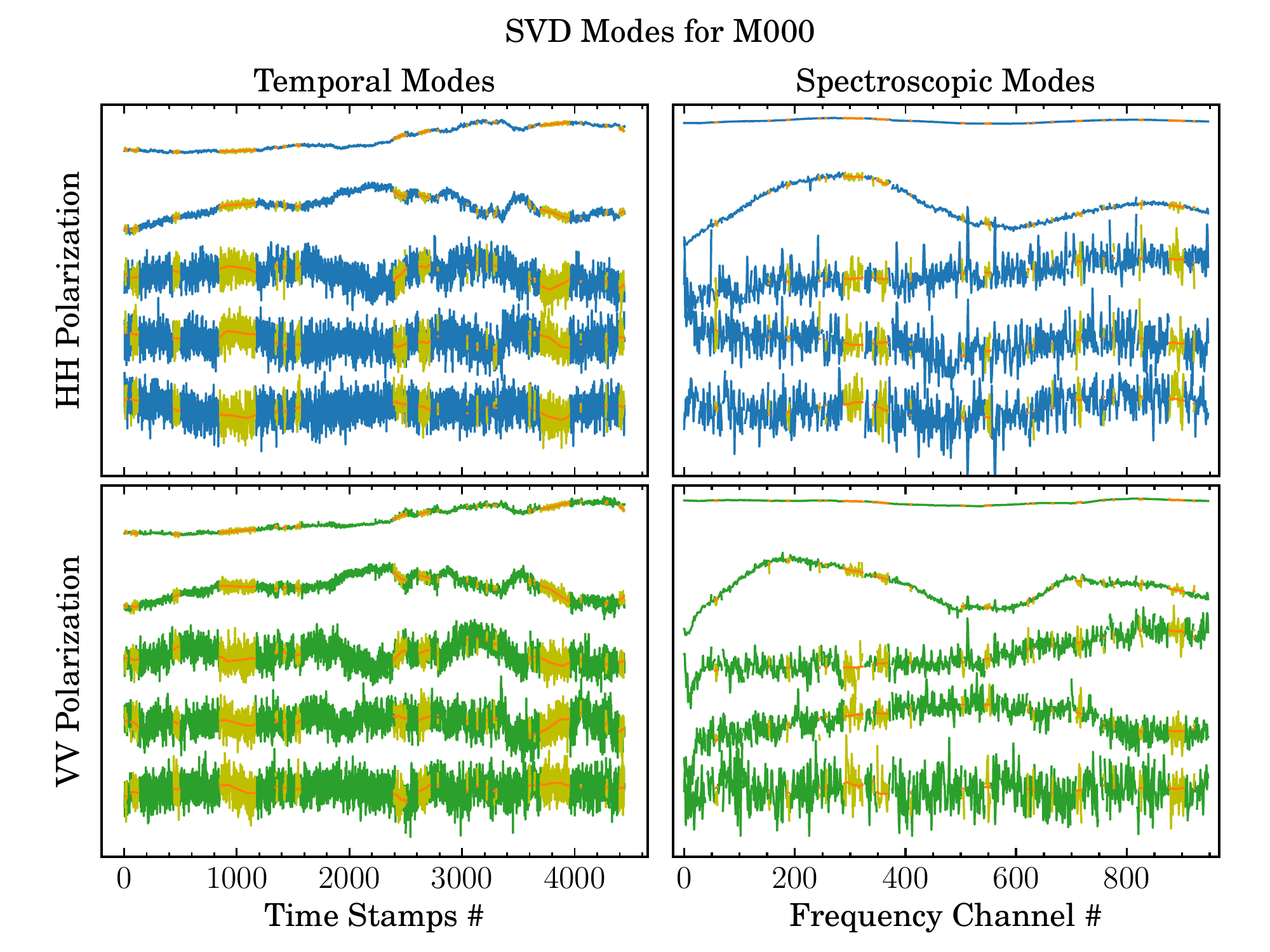}
    \includegraphics[width=0.45\textwidth]{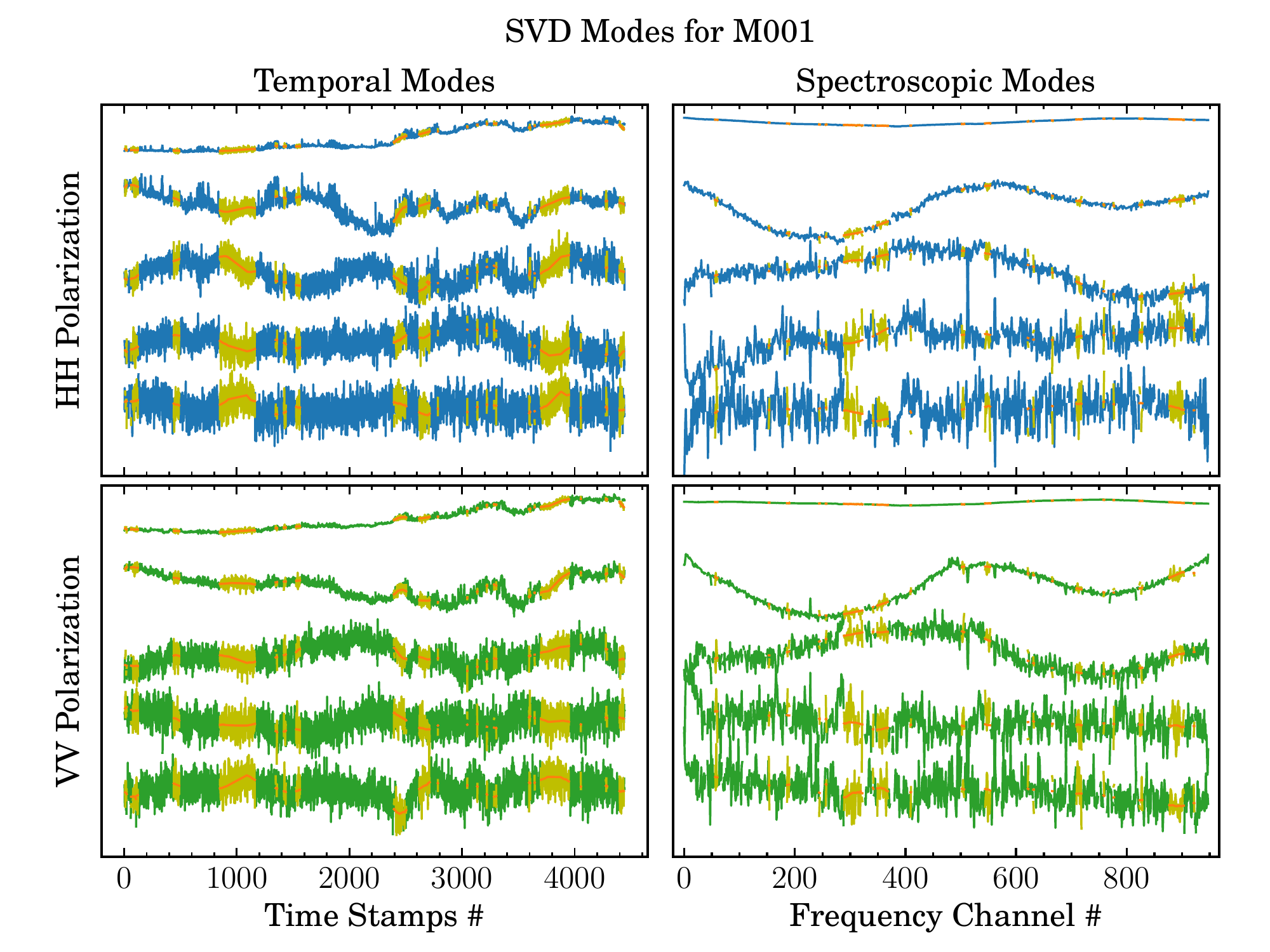}
    \caption{
        Same as \reffg{fig:svdmodes} but shown with the modes of antenna M000 (left)
        and M001 (right) of the SCP19 data.
        The two polarizations are shown in the top and bottom sub-panels;
        and the temporal and spectroscopic modes are shown in the
        left and right sub-panels, respectively.
    }\label{fig:svdmodes19}
\end{figure*}

We apply Singular Value Decomposition (SVD) to the time-ordered data in order to extract its strongest components. If the 1/f noise is strongly correlated in frequency, we expect the SVD will be able to remove it while keeping the 21cm signal unaffected on the scales of interest. Foregrounds should also be removed in this process. SVD corresponds to the following matrix decomposition:
\begin{align}
\label{eq:SVD}
    {\bf D} = {\bf U}{\bf \Lambda}{\bf V}^{\dagger},
\end{align}
where ${\bf D}$ is the data matrix with shape of $n_{t} \times n_{\nu}$;
the columns of ${\bf U} = \{u_0, ..., u_{n_t}\}$ and ${\bf V} = \{v_0, ..., v_{n_{\nu}} \}$
are the temporal and spectroscopic modes and 
${\bf \Lambda}$ is a $n_{t} \times n_{\nu}$ diagonal singular value matrix. The symbol $\dagger$ denotes the conjugate transpose.
Both the temporal and spectroscopic
modes are sorted according to their singular values, and the first $N_{\rm m}$
modes are subtracted,
\begin{align}
    {\bf D}^{\rm c} = {\bf D} - {\bf U} {\bf \Lambda}' {\bf V}^{\dagger},
\end{align}
where ${\bf \Lambda}'$ equals ${\bf \Lambda}$ with diagonals beyond $N_{\rm m}$
set to $0$. It can be further expressed as,
\begin{align}
    {\bf D}^{\rm c}
    &= {\bf D} - \sum_{i=0}^{N_{\rm m}} u_i (u_i^{\dagger} {\bf D}) \\
    &= {\bf D} - \sum_{i=0}^{N_{\rm m}} ({\bf D} v_i) v_i^{\dagger},
\end{align}
where $u_i$, $u_i^{\dagger}$, $v_i$ and $v_i^{\dagger}$ are the elements of ${\bf U}$, ${\bf U}^{\dagger}$, ${\bf V}$ and ${\bf V}^{\dagger}$ respectively.

The singular values of the SCP datasets are shown in \reffg{fig:svdvalue}, normalized with the first (largest) 
singular value. The first $5$ singular values are shown on the left of 
the black dashed line and the rest are shown on the right. It is clear that after the first 5 modes there is little variation in the amplitude. Therefore, we restrict the analysis in this paper to the first 5 modes.
The temporal and spectroscopic modes for SCP16 and SCP19 are shown in
\reffg{fig:svdmodes} and \reffg{fig:svdmodes19}, respectively. 
In each panel, the two polarizations are shown in the upper/lower
sub-panels with blue/green colors and the temporal/spectroscopic modes 
shown in the left/right sub-panels, respectively. 
The areas shown in yellow are the re-filled values due to the RFI flagging,
and the orange curves are the Wiener filtered smooth curves.
We will discuss our mask-filling strategy in \refsc{sc:fm}.

The $20\,{\rm Hz}$ and $1\,{\rm Hz}$ data of SCP16 have significant differences
in the singular values, as well as the singular modes.
The left panel of \reffg{fig:svdvalue}
shows the singular values of all three antennas of
both the $20\,{\rm Hz}$ and $1\,{\rm Hz}$ data of SCP16; 
the singular values of the $1\,{\rm Hz}$
data decrease more quickly than those of the $20\,{\rm Hz}$ data. This is 
because the $20\,{\rm Hz}$ data has much shorter observation time than 
the $1\,{\rm Hz}$ data and the SVD modes are dominated by the system noise.
This difference can also be seen with the SVD modes in \reffg{fig:svdmodes}.
The two panels of \reffg{fig:svdmodes} show the SVD modes of the antenna
M017 of SCP16 data, where the $1\,{\rm Hz}$ data are on the left and the $20\,{\rm Hz}$
data are on the right.  We can see that, at least for the first $3$ modes,
both for the temporal and spectroscopic modes, the $1\,{\rm Hz}$ data have
clear variation shapes; but the modes of $20\,{\rm Hz}$ data are mostly noise
dominated.

The singular values of the data of SCP19 are shown in the right panels of
\reffg{fig:svdvalue}. Only the first $6$ antennas are shown here as examples;
the rest of the antennas have the same trend as these first $6$ antennas.
The SVD modes of the SCP19 data are shown in the \reffg{fig:svdmodes19},
in which the left and right panel show the modes of two different 
antennas as examples. Similar to the SCP16 $1\,{\rm Hz}$ data, with long enough 
observation time, the system variations both along time and frequency are
well represented with the first several singular modes.

\section{Power spectrum in the presence of RFI flagging}\label{sc:fm}

The gaps due to the RFI flagging result in significant window function effects 
in the final power spectrum. 
To reduce the impact of this effect
in our analysis, we fill the missing data
with values reconstructed with the SVD modes. The mask-filling strategy 
is described below.

First of all, some data are removed across either whole frequency channels or
whole time samples. Frequency channels that are contaminated by some
known narrow band RFI are fully masked for the whole observation time.
The \hi emission line of the Milky Way, which is in our selected frequency
band, is also removed. On the other hand, some short time duration occasional RFI, 
for example due to transiting satellites, are masked across the whole frequency band.
By ignoring the masked frequency channels and time stamps, the rest of the data
are merged into a continuous frequency-time matrix. SVD is applied
to this merged data.

We take the resulting first $5$ temporal and spectroscopic modes, which have larger
singular values than the noise modes, and fill the masked regions with linear 
interpolation. The interpolation is applied for each of the temporal and 
spectroscopic modes individually. However, if the singular modes are noise
dominated, such as the first several modes of SCP16 $20\,{\rm Hz}$ data, the 
interpolation is ignored and the masked region is filled with the mean
value of the singular mode. Then we apply a Wiener filter to each of the
masked-filled singular modes to make the filled values smoothly connect with
the unmasked region. The smoothed filling values are shown with orange curves 
in \reffg{fig:svdmodes} and \reffg{fig:svdmodes19}.

\begin{figure*}
    \small
    \centering
    \begin{minipage}[c]{0.70\textwidth}
        \centering
        \includegraphics[width=0.49\textwidth]{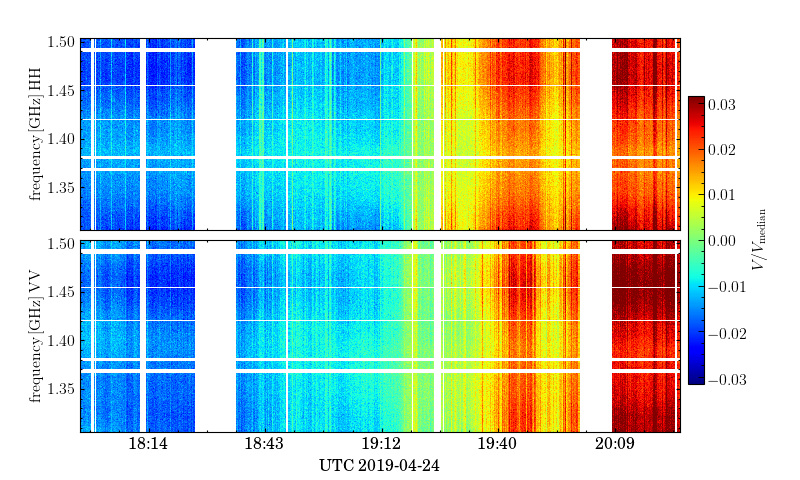} 
        \includegraphics[width=0.49\textwidth]{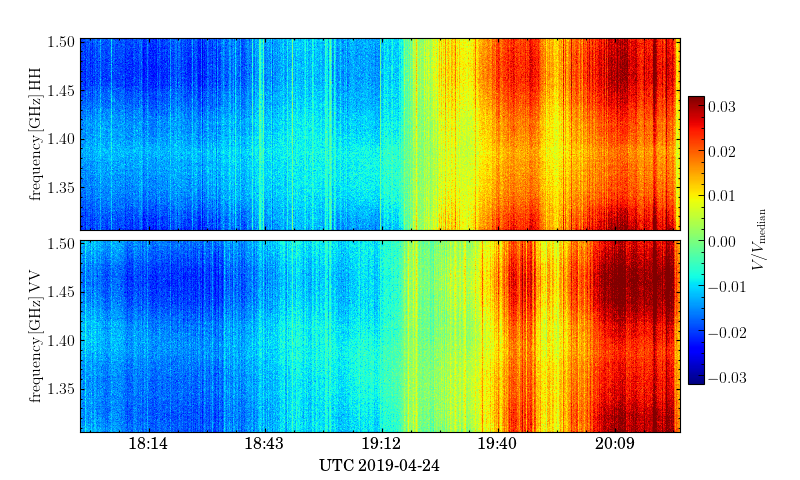} 
        \includegraphics[width=0.49\textwidth]{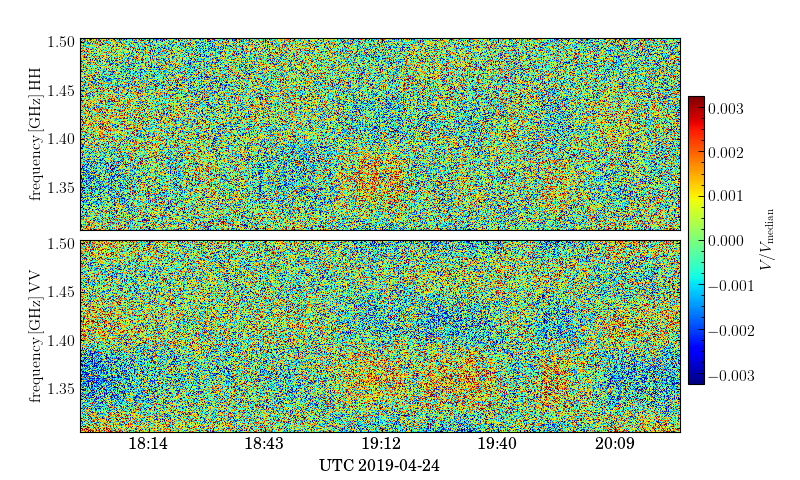}
        \includegraphics[width=0.49\textwidth]{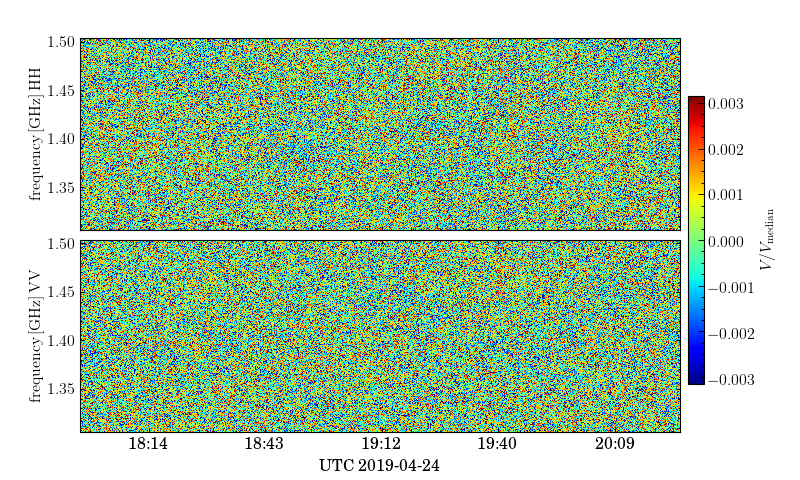}
        \includegraphics[width=0.49\textwidth]{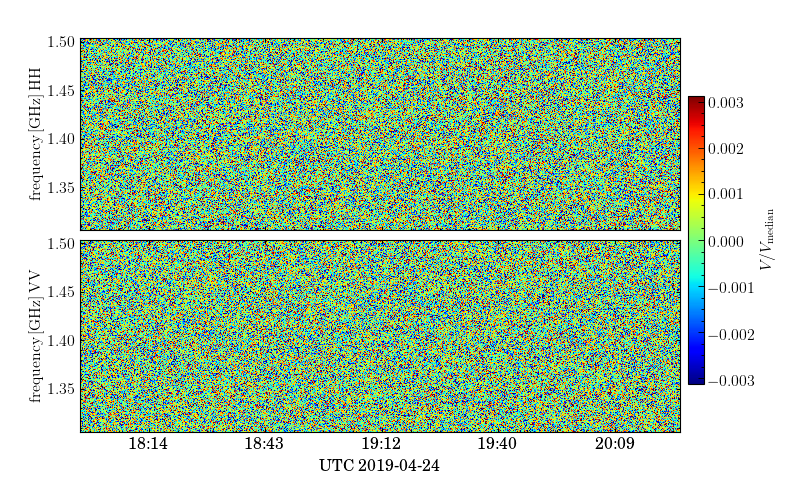}
        \subcaption{}\label{fig:wfsvd}
    \end{minipage}
    \hfill
    \begin{minipage}[c]{0.29\textwidth}
        \centering
        \includegraphics[width=0.75\textwidth]{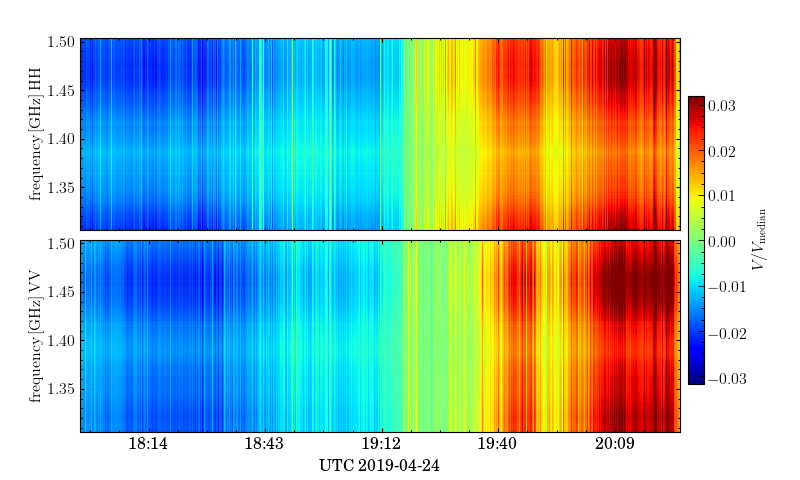}\vspace{-0.25cm}
        \includegraphics[width=0.75\textwidth]{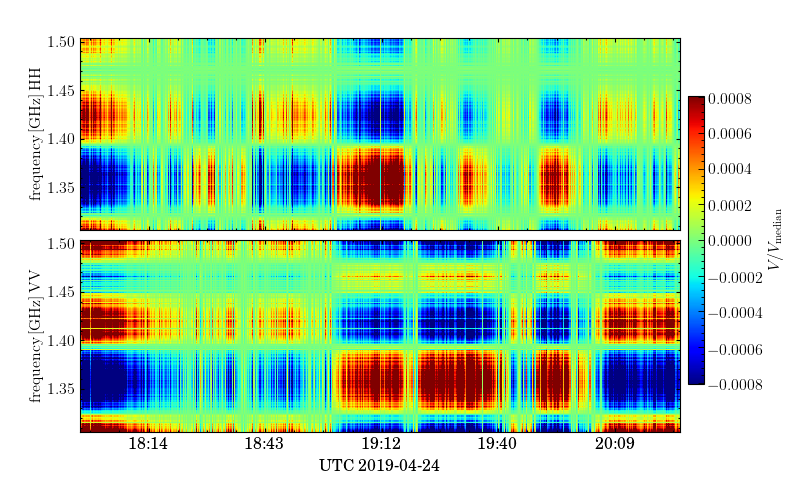}\vspace{-0.25cm}
        \includegraphics[width=0.75\textwidth]{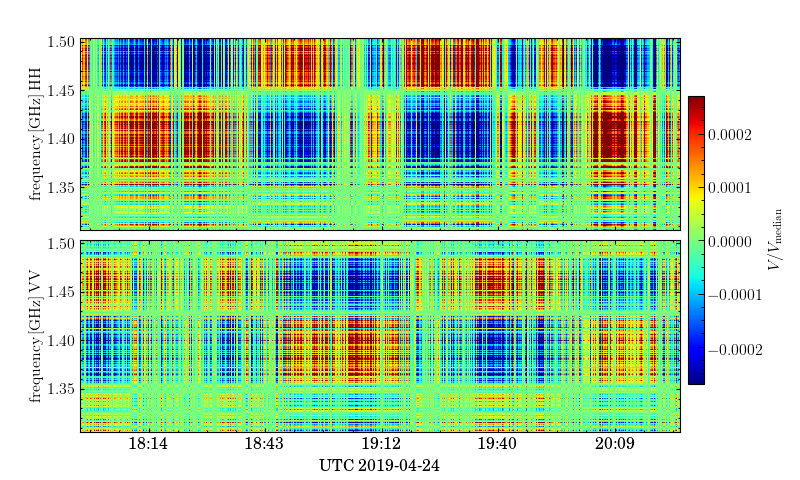}\vspace{-0.25cm}
        \includegraphics[width=0.75\textwidth]{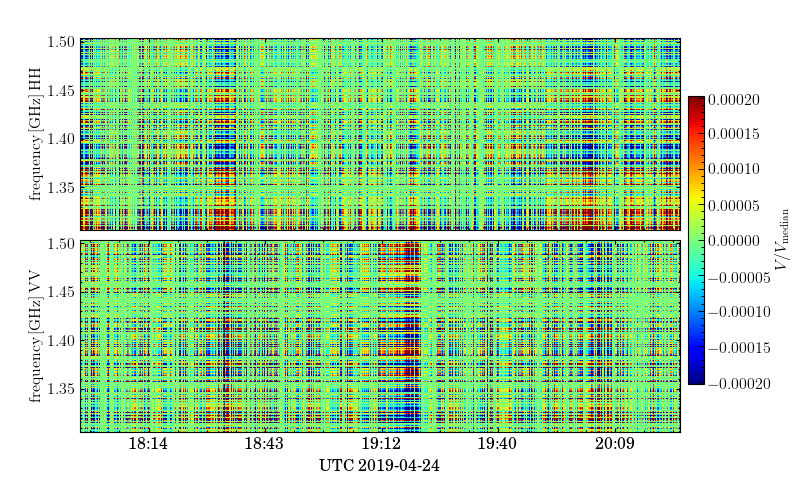}\vspace{-0.25cm}
        \includegraphics[width=0.75\textwidth]{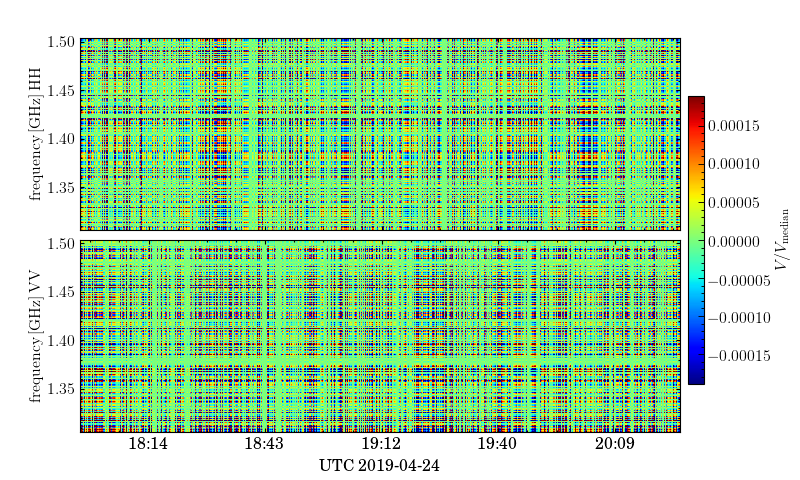}
        \subcaption{}\label{fig:wfmod}
    \end{minipage}
    \caption{
        (a):
            The top-left panel shows the waterfall plots of the mean-subtracted
            data before SVD mode subtraction; The top-right panel shows the same
            data with the RFI masks filled; The middle-left, middle-right and bottom
            panels are the waterfall plots of the mask filled data with $1$, $2$ and $5$
            SVD modes subtracted. 
        (b):
            The waterfall plots of the first $5$ SVD modes from the mask filled data.
        All these data are the SCP19 observation with antenna M001.
    }
\end{figure*}

We then subtract the Wiener-filter-smoothed singular modes from the 5 modes, estimate the
r.m.s. of the residuals for each mode and add random noise to the filling values
according to the residual r.m.s. of each mode. The noise-added filling
values are shown in yellow in \reffg{fig:svdmodes} and \reffg{fig:svdmodes19}.
Using only these 5 modes we construct a new dataset (through \refeq{eq:SVD}) which now has values in the missing gaps. 
We could be tempted to use these values to fill the flagged gaps in our original data.
However, this reconstructed data with the first $5$ modes still has noise missing (from the remaining modes).
In order to account for this, we subtract this new dataset from the original dataset and estimate an overall noise r.m.s. using the non-flagged part of the data. We then add random noise to the new dataset using this r.m.s and use its values to fill the flagged gaps in the original data.
The original masked data and masked filled data of SCP19 (antenna M001)
are shown in the top-left and top-right panels of \reffg{fig:wfsvd}, respectively.

Finally, we perform SVD a second time to the mask-filled data.
The waterfall plots with the first $1$, $2$ and $5$ modes removed are shown 
in the middle-left, middle-right and bottom panels of \reffg{fig:wfsvd}.
The corresponding first $5$ modes are shown in \reffg{fig:wfmod}.
The power spectrum estimation discussed in \refsc{sec:results}
is performed with the mask-filled data.

\section{Results and Discussion}\label{sec:results}

\begin{figure*}
    \small
    \centering
    \begin{minipage}[t]{0.49\textwidth}
        \centering
    \includegraphics[width=\textwidth]{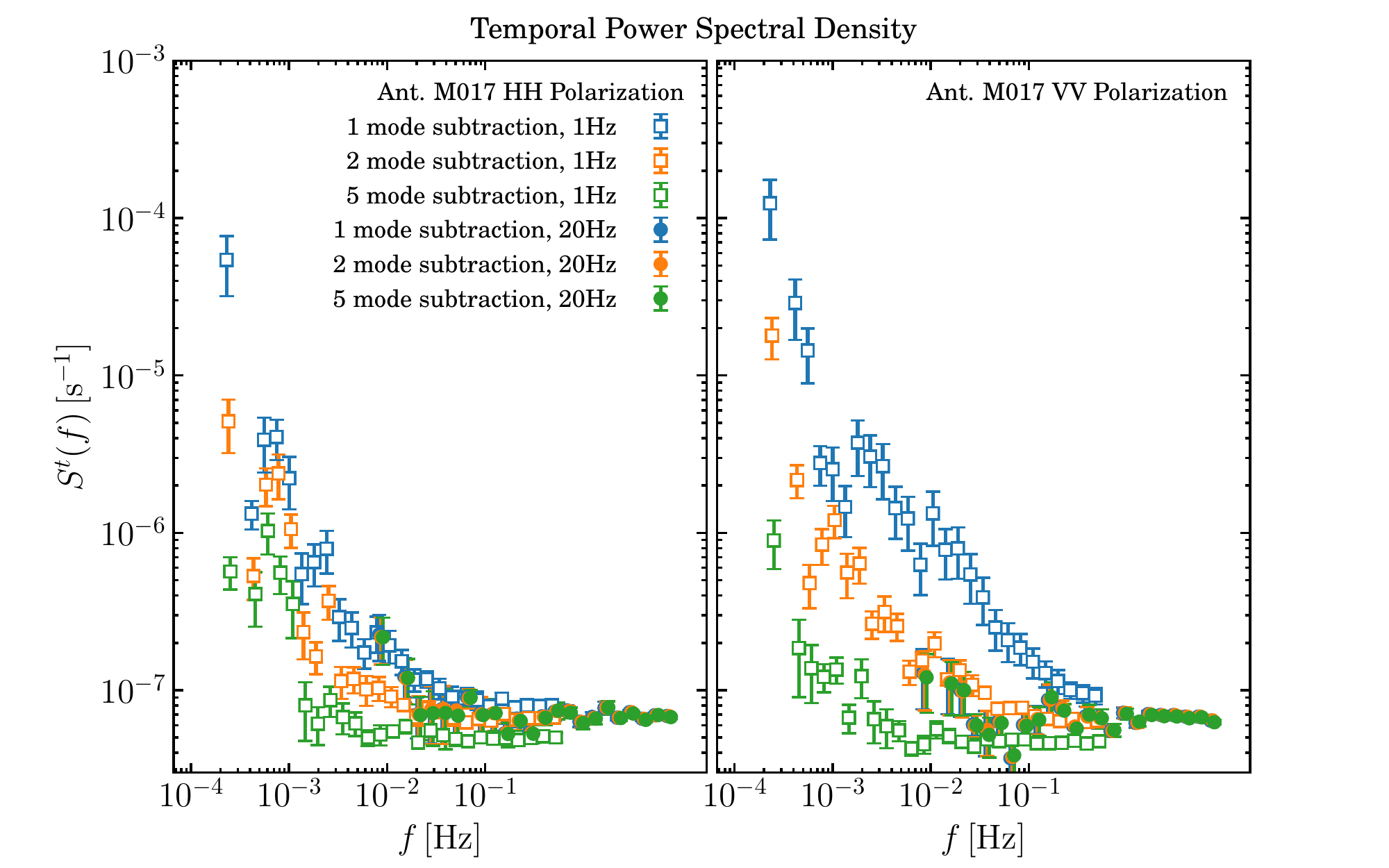}\\
    \includegraphics[width=\textwidth]{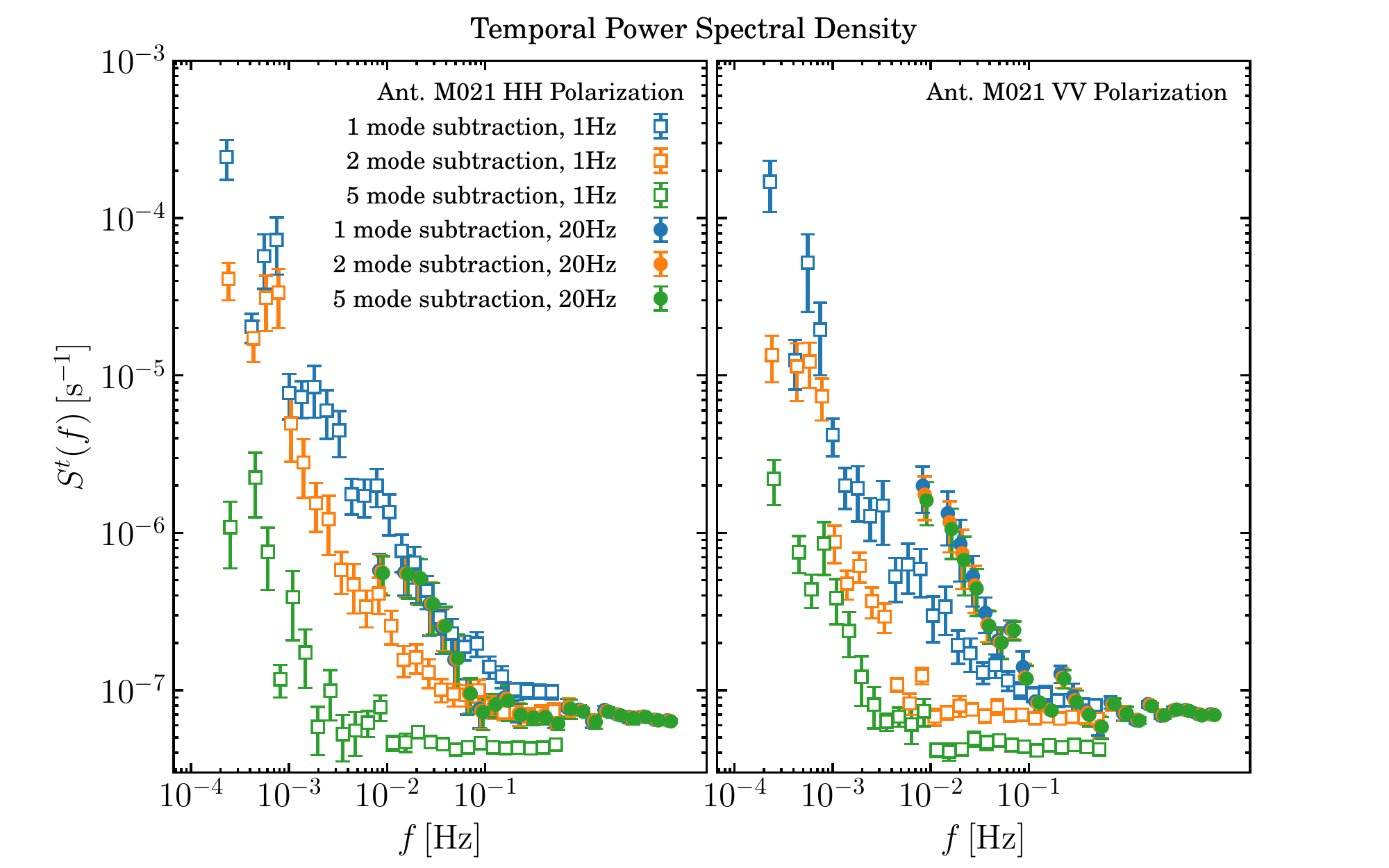}\\
    \includegraphics[width=\textwidth]{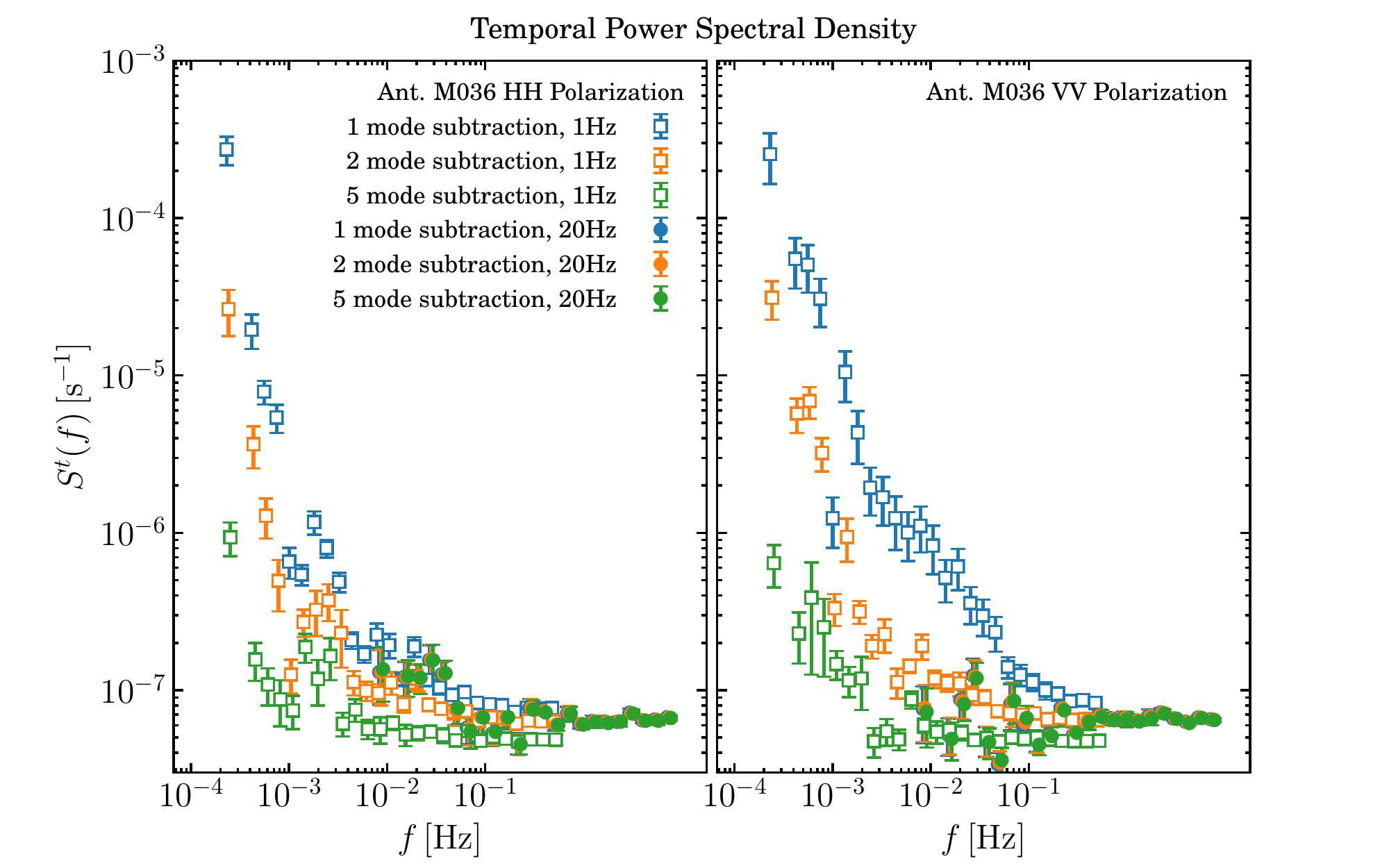}
    \caption{
        The temporal power spectrum density of the SCP16 data.
        The results of the $3$ antennas are shown in different panels.
        In each panel, the results of two polarizations are shown in the left
        and right subpanels; 
        The results with $1$, $2$ and $5$ mode subtraction
        are shown in different colors as labeled in the legend.
        The errors of the power spectrum density are estimated via the variance
        over different frequency channels.
    }\label{fig:results16}
    \end{minipage}
    \hfill
    \begin{minipage}[t]{0.49\textwidth}
        \centering
    \includegraphics[width=\textwidth]{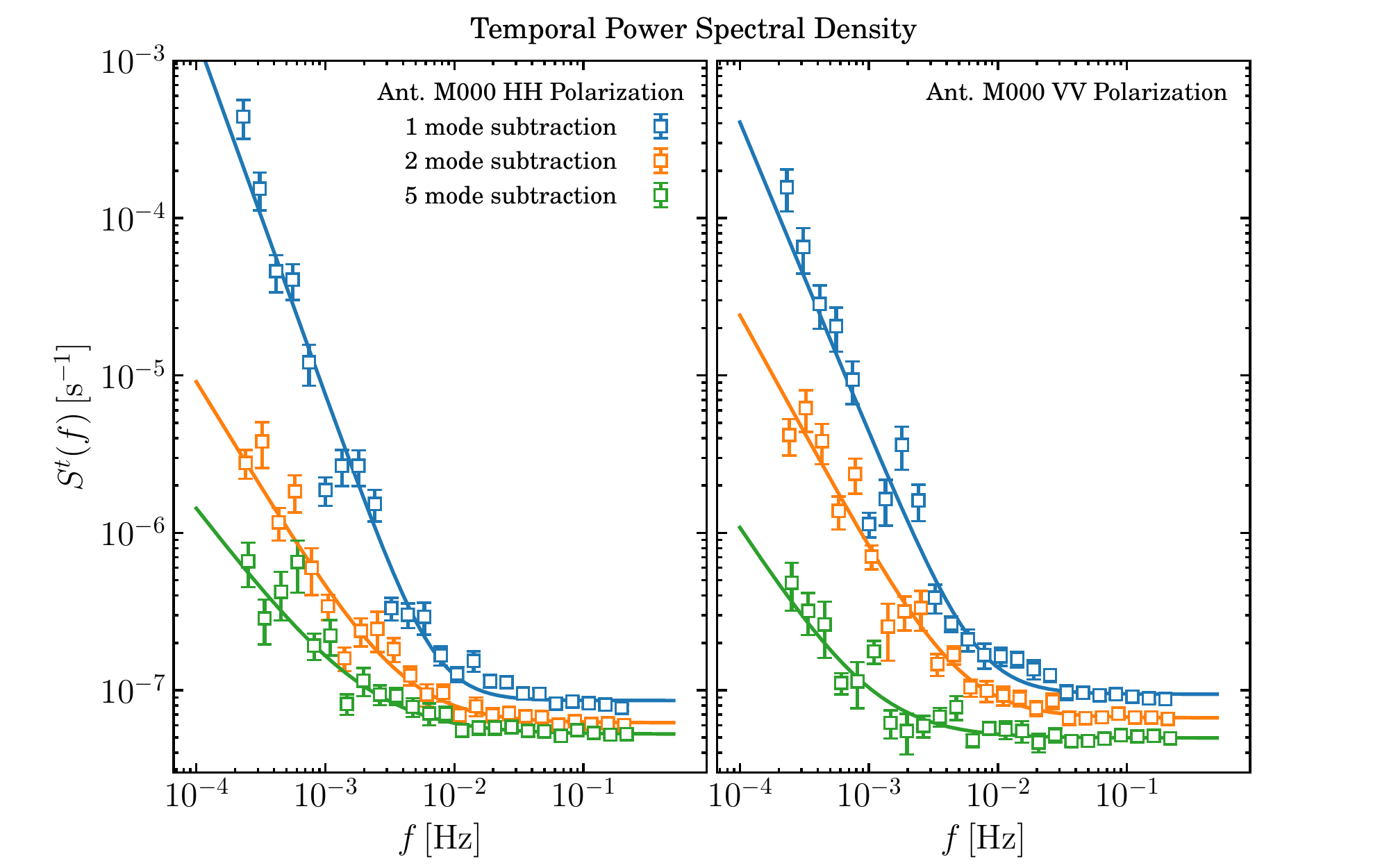}\\
    \includegraphics[width=\textwidth]{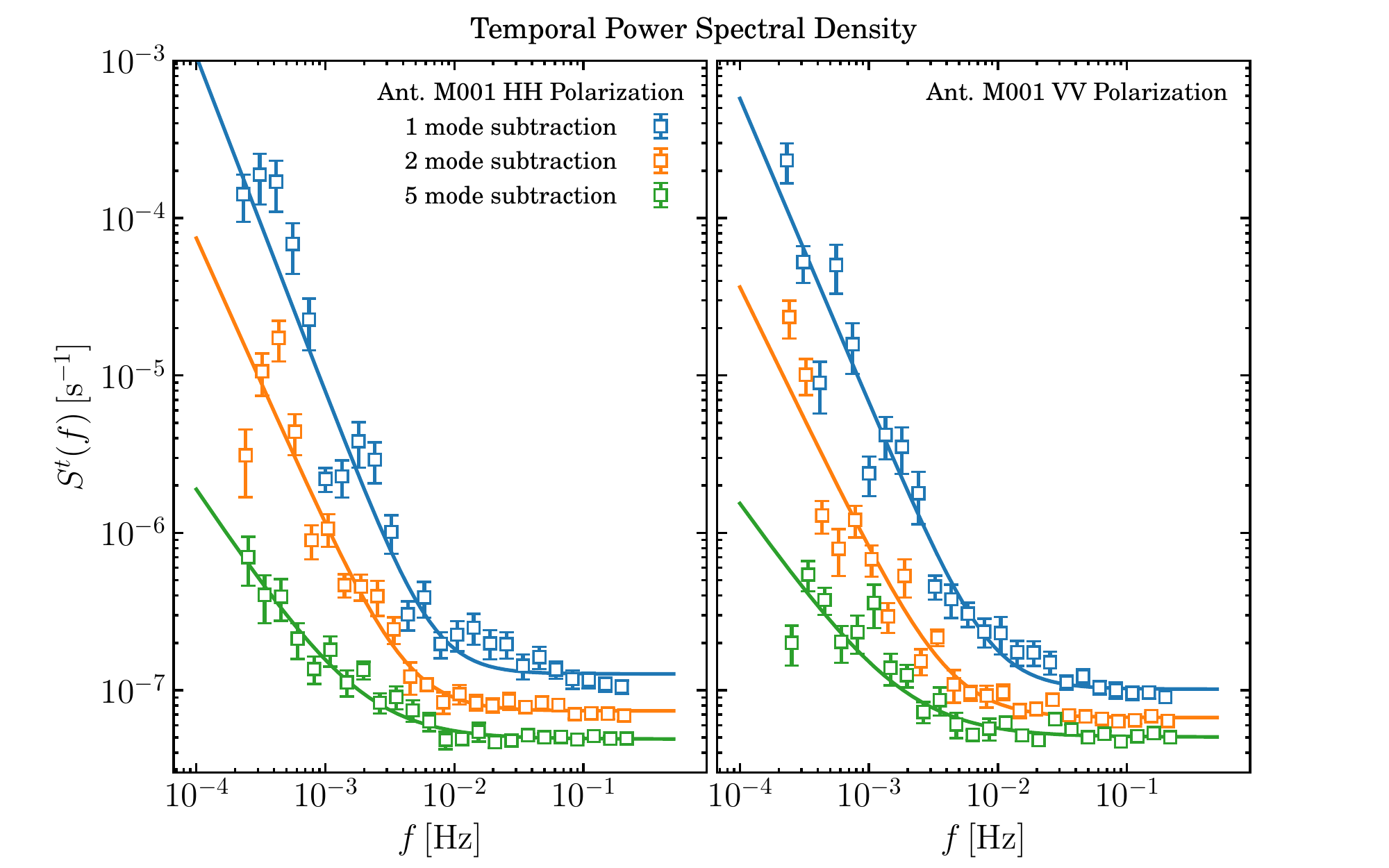}\\
    \includegraphics[width=\textwidth]{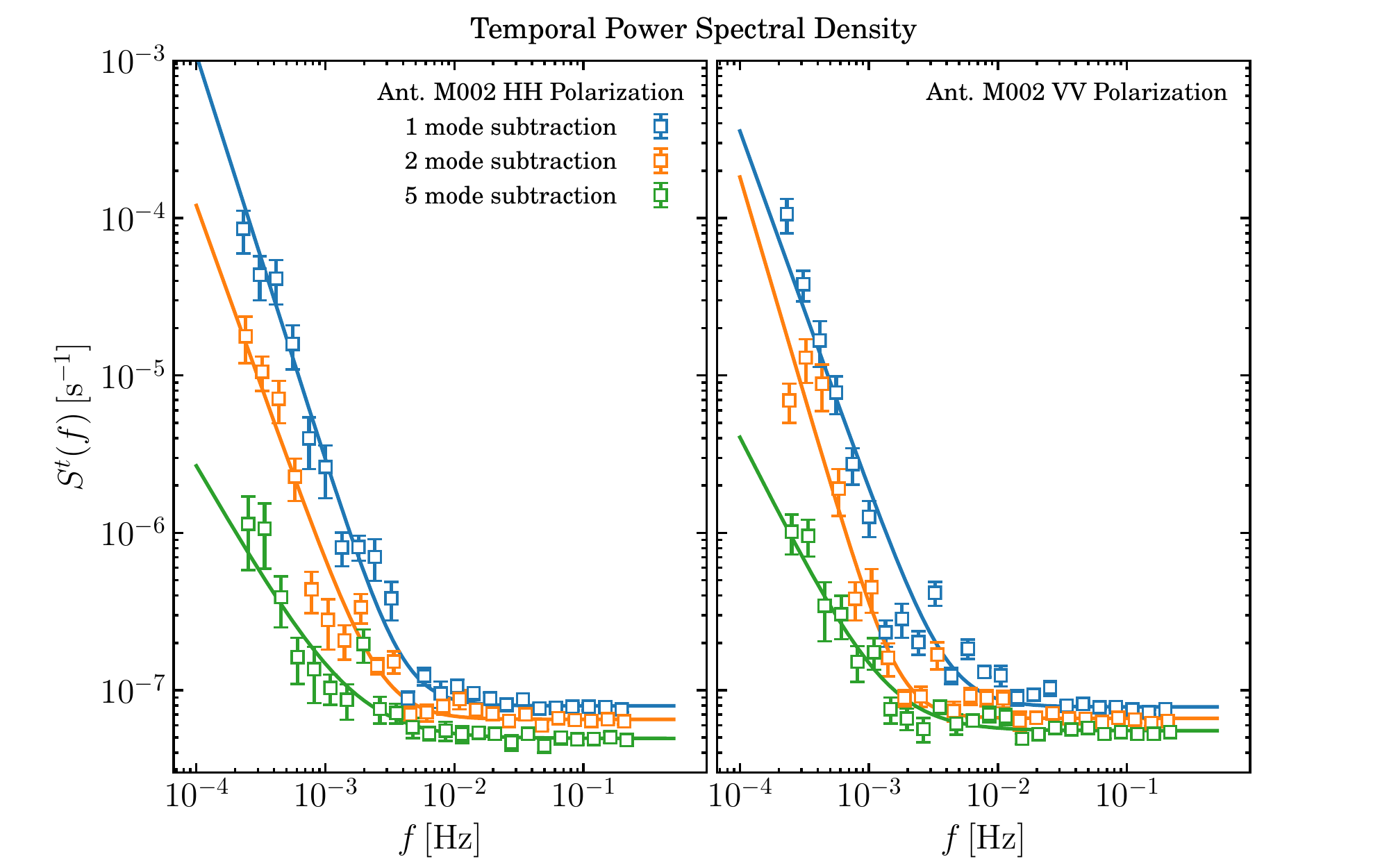}
    \caption{
        Same as \reffg{fig:results16} but for SCP19 datasets.
        The results of the first $3$ antennas are shown in different panels.
        The solid lines are the fitted 1/f noise temporal model using 
        \refeq{eq:tmodel}.
    }\label{fg:tps19}
    \end{minipage}
\end{figure*}

\subsection{The Temporal Power Spectrum Density}\label{sec:tpsd}

We now focus on the analysis of the power spectrum along the time domain and how it compares to our model.
The temporal power spectrum is estimated by Fourier transforming the data along the 
time axis. 
Before the power spectrum estimation, we reduce the frequency resolution down 
to $\sim20\,{\rm MHz}$ by averaging every 
$100$ frequency channels. The frequency averaging can reduce the white noise
level and shift the knee frequency to the high-end of $f$.
However, as we discussed before, the shift of the knee frequency is also dependent 
on the frequency correlation of the 1/f noise.

The temporal power spectrum density results of SCP16 data observed with the 
three antennas are shown in \reffg{fig:results16}, and the results of the 
first $3$ antennas of SCP19 data are shown in \reffg{fg:tps19}. 
The two polarizations are shown in the left and right subpanels. 
We do not show the raw data as it is mostly dominated by external sources (e.g. sky and ground pick up). This should be very smooth in frequency with our observation and most of it should be removed with the first SVD mode.
The results with $1$, $2$ and $5$ mode subtraction
are shown in different colors as labeled in the legend.
The errors of the power spectrum density are estimated via the variance
over different frequency channels.
Significant 1/f-type noise in the power spectrum is visible in all the plots.
After SVD modes subtraction, the 1/f-type noise power is reduced and 
flat white noise dominates the power spectrum over a wide frequency range,
with a clear knee frequency visible. This indicates that most of the 1/f noise is correlated in frequency.

For the SCP16 data,
the results with $1\,$Hz sampled data are shown with empty markers and 
the $20\,$Hz sampled data with filled markers. Both show good agreement of the white
noise floor (\reffg{fig:results16}) . This level is also consistent with the 
theoretical value in \refeq{eq:tmodel} given by $1/\delta\nu$, which 
at $\delta\nu=20\,{\rm MHz}$ corresponds to $\sim 5 \times 10^{-8}$.
The noise floor is also slightly reduced with the $5$ modes subtraction. Note however that it is still equal or above the predicted theoretical white noise value. Mostly likely, this noise floor reduction is due to the removal of correlated modes in frequency but that are fluctuating on short time-scales.

We also notice that the SVD mode subtraction does not work well with antenna 
M021 for the $20\,$Hz sampled data, especially for the VV polarization. 
As shown in \reffg{fig:svdvalue}, the singular values of this antenna with $20\,$Hz 
sampled data are barely reduced. With $1\,$Hz sampled data, 
M021 has larger singular values for the second and third modes 
compared with the other two antennas. By looking at the waterfall plot in
\reffg{fig:wf}, M021 has more fluctuations than the other two antennas. 
This indicates that the system of M021 was quite unstable during the 
observation of SCP16.

The SVD mode subtraction works well for the SCP19 data, as shown in \reffg{fg:tps19}.
We only show the results of the first $3$ antennas because the other antennas have 
similar behaviors. Comparing to SCP16 datasets, the $f$ upper bond of SCP19 is
limited by the lower time sampling resolution. 
The solid lines in \reffg{fg:tps19} are the fitted 1/f noise temporal model using 
\refeq{eq:tmodel}. Again we do not fit to the raw data here as that would require a more complex model, possibly with a running power law. Once we remove one or more modes the fit using \refeq{eq:tmodel} works quite well. Removing two modes is quite conservative and we expect it to be done for most data analysis. 

After the subtraction of the first singular mode, the 1/f type correlation in time is
highly reduced and the knee frequency is reduced below $10^{-2}\,$Hz. A clear 
white noise floor is shown at the high-$f$ end and the noise floor is 
$\sim 5 \times 10 ^{-8}$, which is consistent with both the SCP16 data, as well as 
the model prediction. 
With additional singular modes subtraction, the knee frequency is further reduced. 
Again, the noise floor is slightly reduced with $5$ mode subtraction.


\subsection{2D Power Spectrum Density}\label{sec:2dpsd}

\begin{figure*}
    \small
    \centering
    \includegraphics[width=0.48\textwidth]{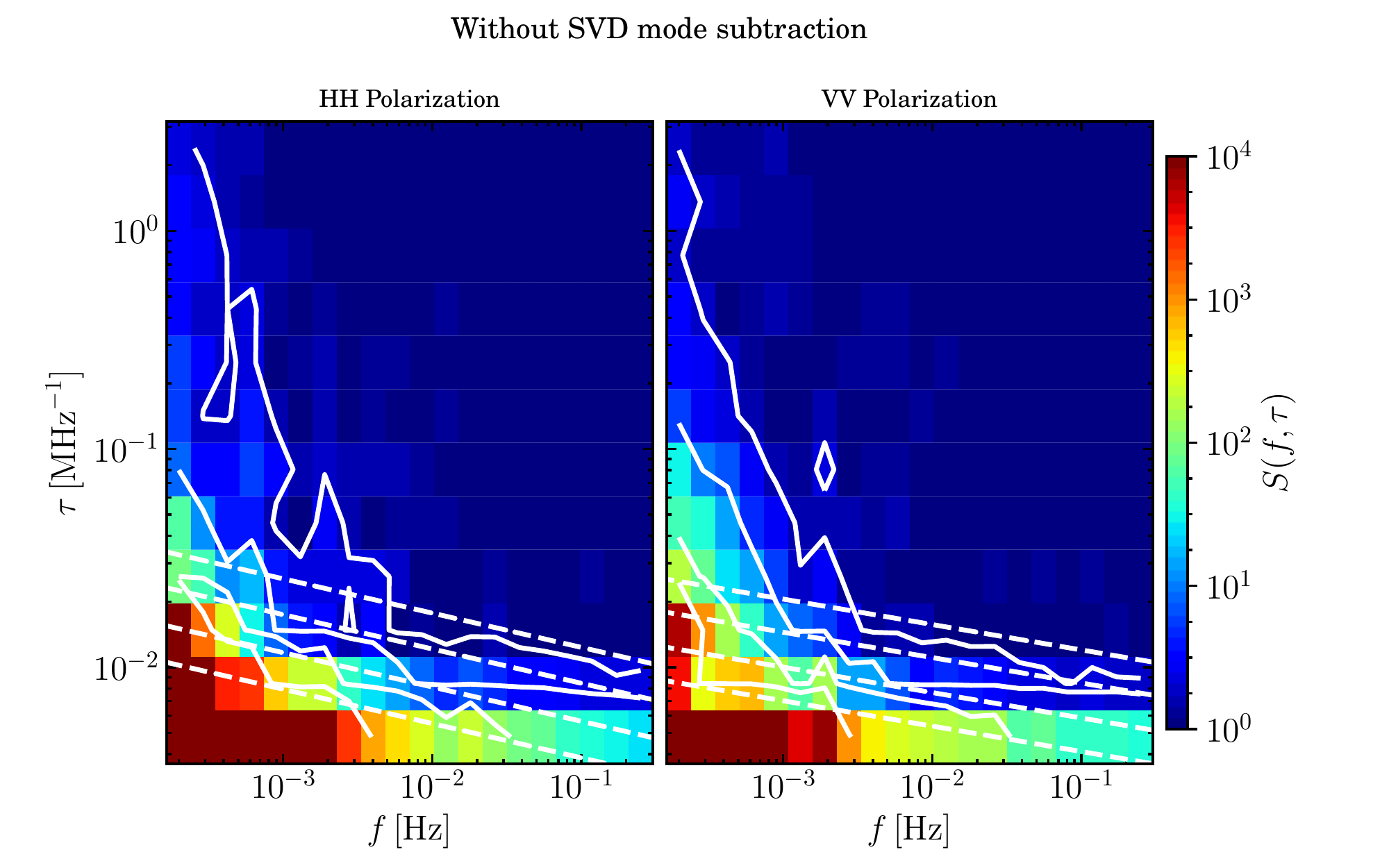}
    \includegraphics[width=0.48\textwidth]{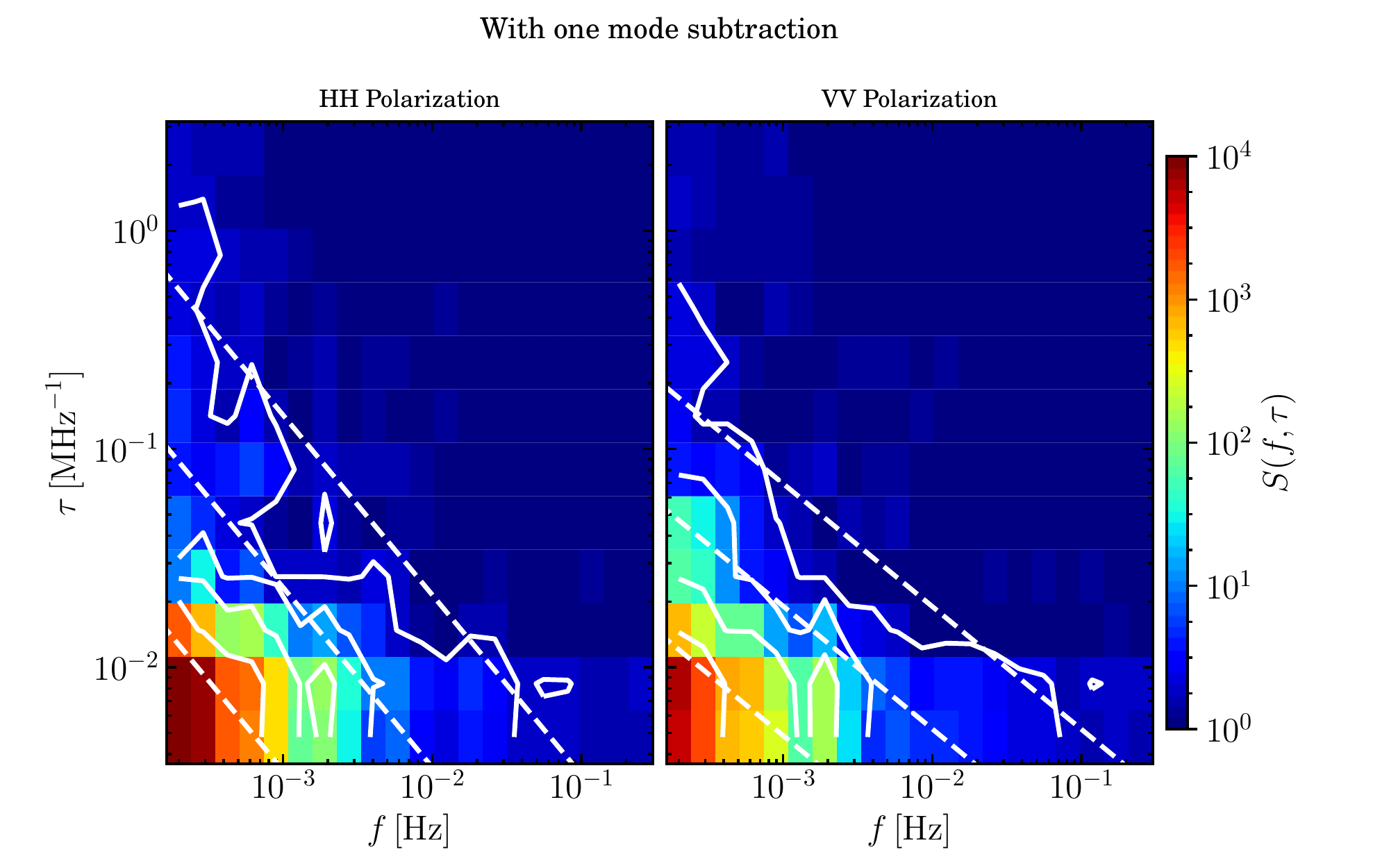}
    \includegraphics[width=0.48\textwidth]{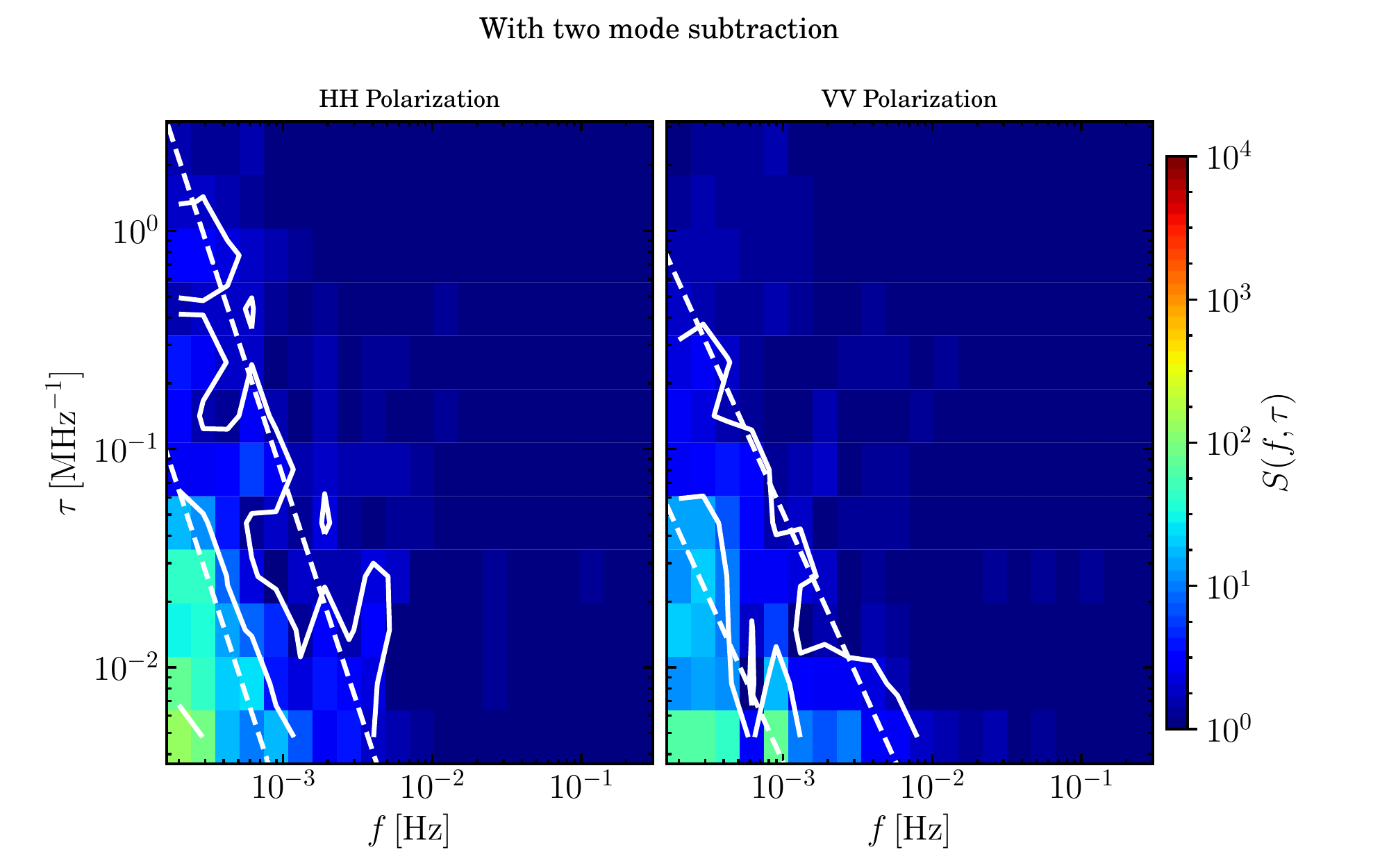}
    \includegraphics[width=0.48\textwidth]{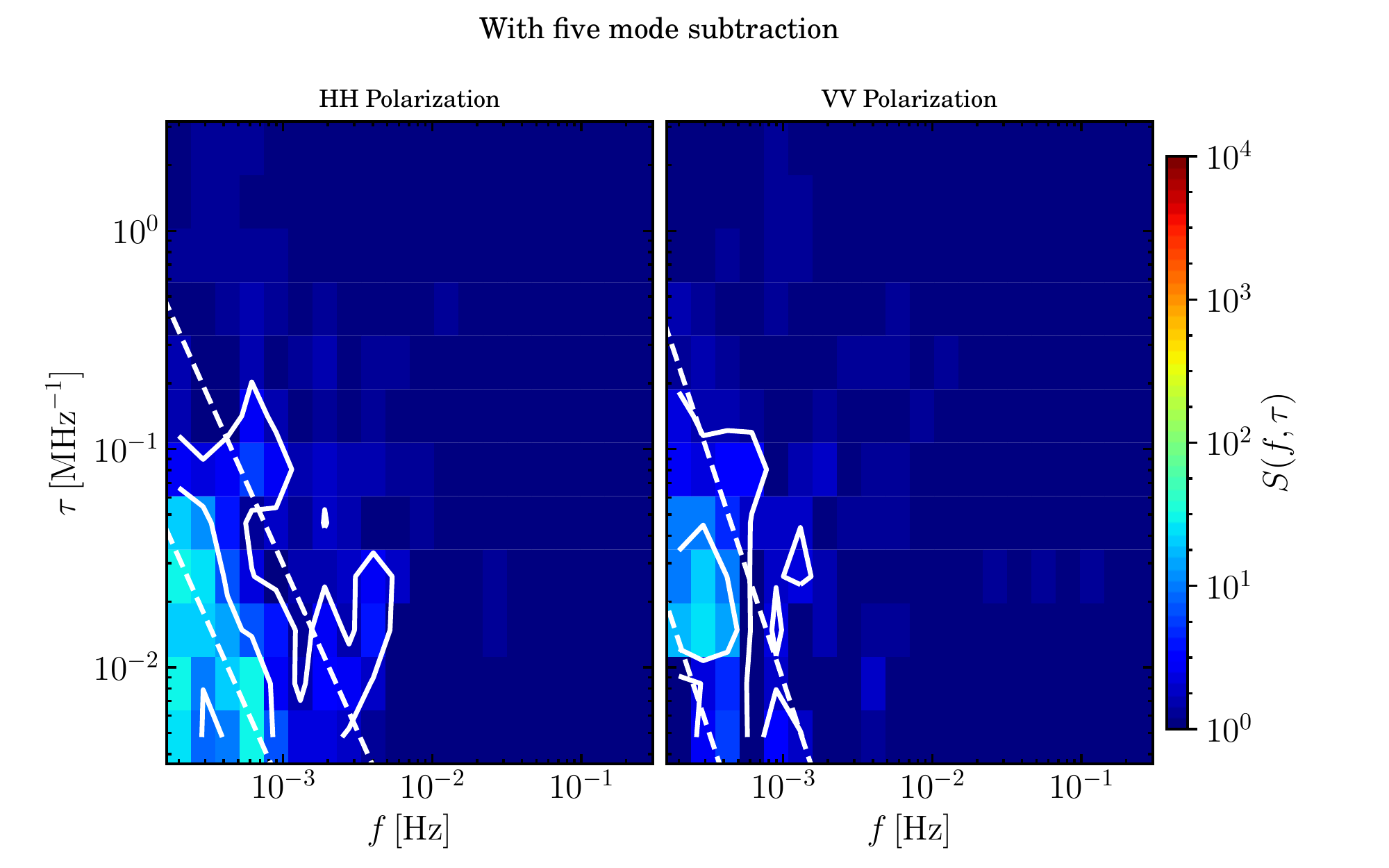}
    \caption{
        The 2D power spectrum density estimated with SCP19 data.
        From the top-left to the bottom-right, each panel shows the 
        result with no SVD modes subtraction, one mode, two modes and
        five modes subtraction. The two polarization are shown in the
        left/right sub-panels. All the plots are truncated with the same
        color scale. 
        The white contours shows the levels $2$, $10$, $10^2$, and $10^3$. 
        and the dashed contours show the fitted power spectrum model at the same levels. 
    }\label{fg:ps2d19}
\end{figure*}

\begin{figure*}
    \small
    \centering
    \includegraphics[width=0.8\textwidth]{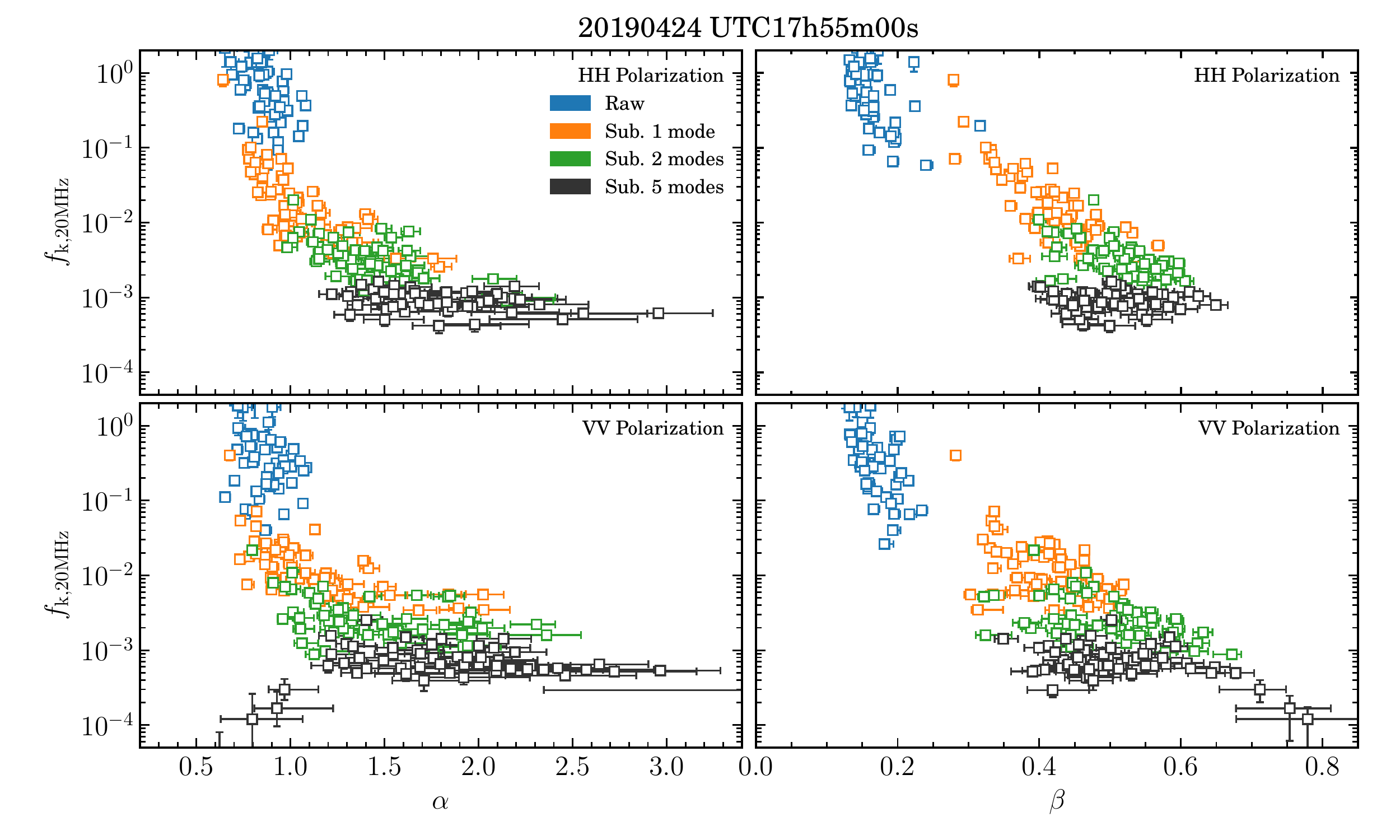}
    \caption{
        The fitted $f_{\rm k, 20MHz}$ versus $\alpha$ (left panels)
        and $\beta$ (right panels) for each of the antennas.
        Top/bottom panels show the results of HH/VV polarizations. 
        The results with different number of modes subtracted are
        shown with different colors.
    }\label{fg:fkab}
\end{figure*}

\begin{figure*}
    \small
    \centering
    \includegraphics[width=0.48\textwidth]{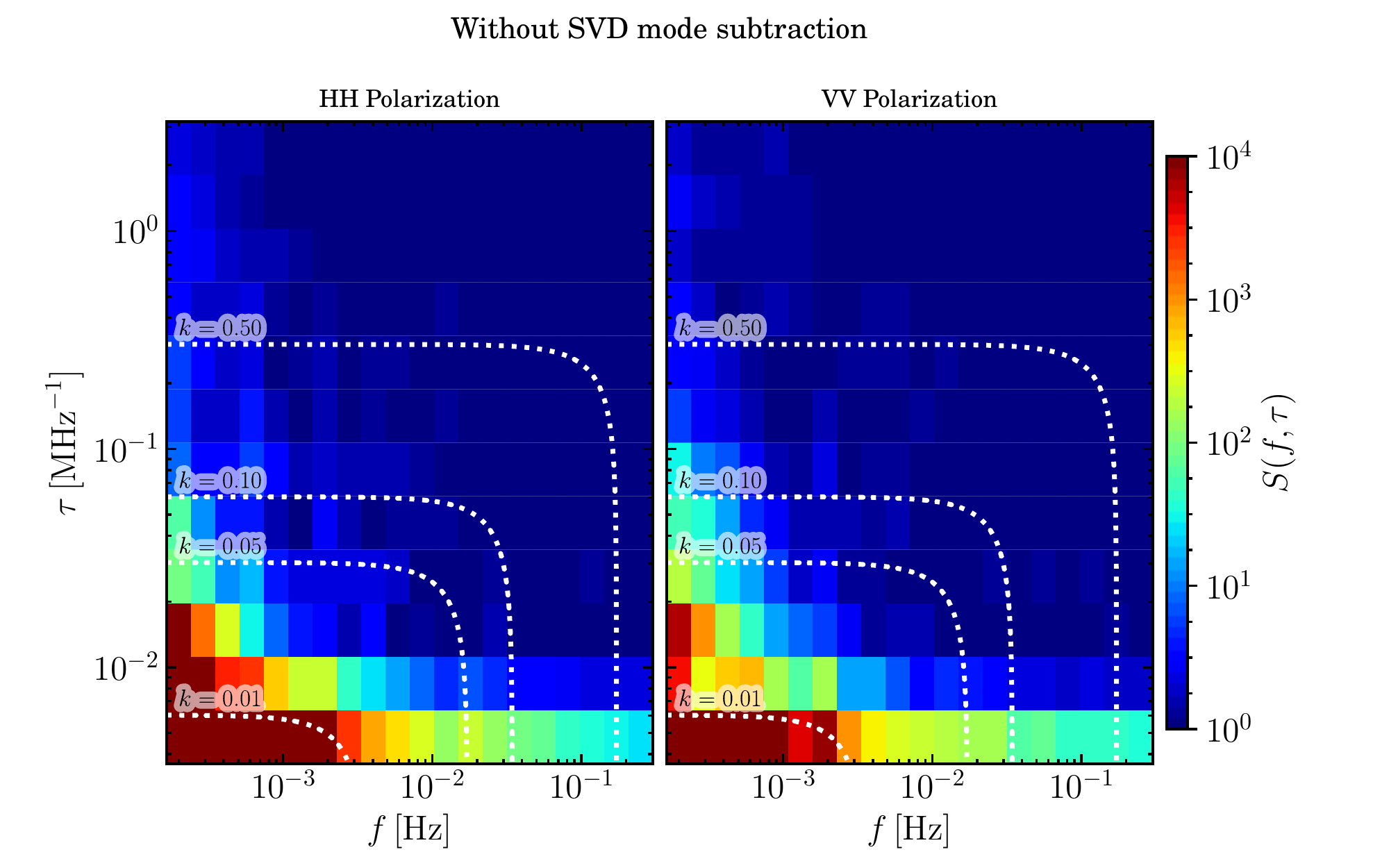}
    \includegraphics[width=0.48\textwidth]{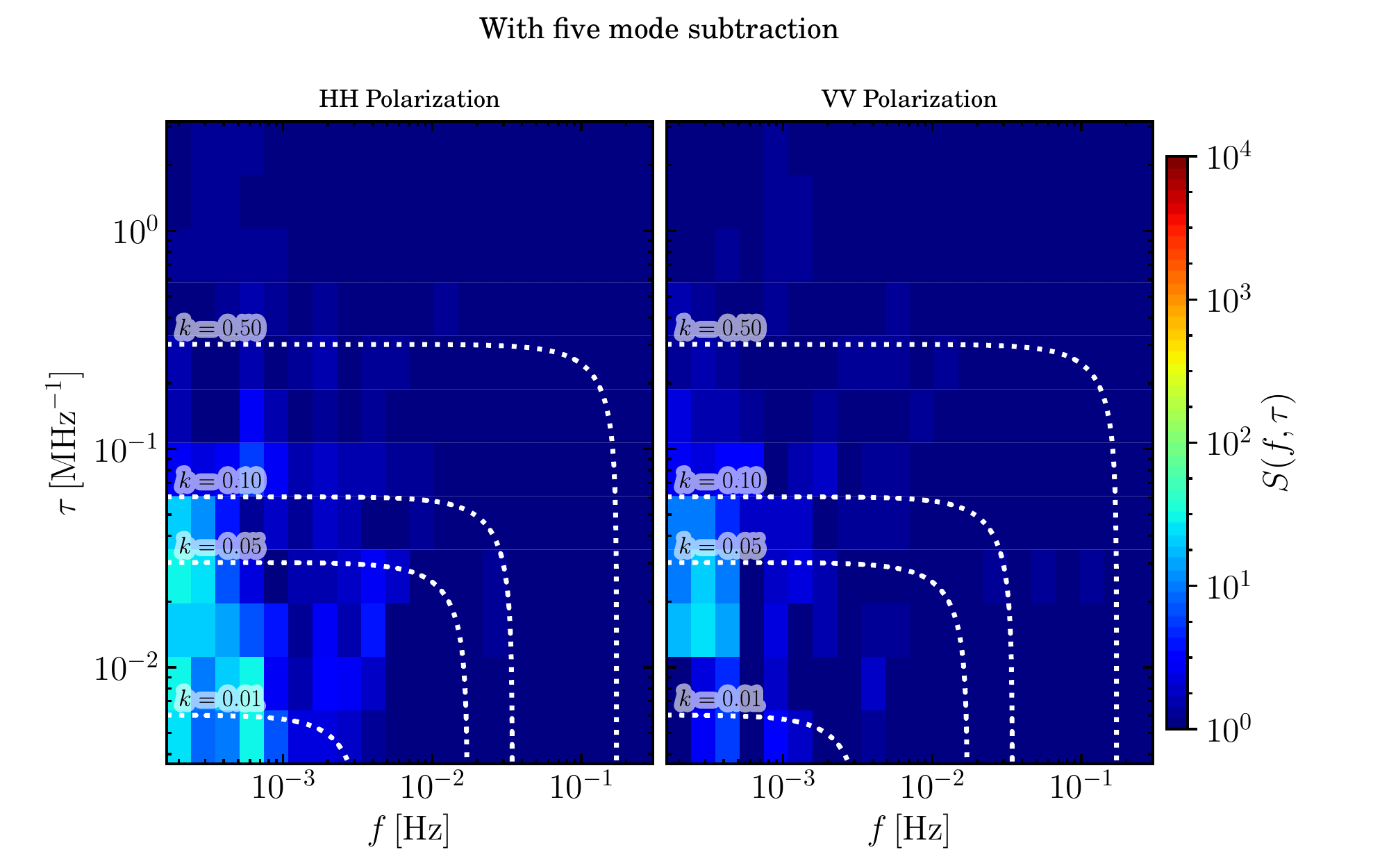}
    \caption{
        Same as \reffg{fg:ps2d19} but only for the results with
        no SVD mode subtraction (left) and five mode subtraction (right). 
        The white dotted lines indicate the corresponding cosmological 
        scales projecting to the $f-\tau$ space, assuming observations at 
        $900\,{\rm MHz}$ with scanning speed of $5\ {\rm arcmin}\,s^{-1}$.
        $k$ is in units of ${\rm Mpc}^{-1}h$.
    }\label{fg:ps2d19wBAO}
\end{figure*}

The 2D power spectrum density is estimated by Fourier transforming the data 
along both the time and frequency axes. The results for SCP19
are shown in \reffg{fg:ps2d19} for one antenna (M000)
as an example. From top-left to the bottom-right panels, it shows the 2D power
spectrum with $0$, $1$, $2$ and $5$ SVD modes subtracted, as labeled in 
the title of each panel. The results for the two polarizations are shown in
the left and right subpanels, respectively. 
The white contour shows the levels 
$2$, $10$, $10^2$ and $10^3$. The dashed contours show the fitted 2D 
power spectrum model at the same power spectrum levels as the measurements.
Since we are analysing the power spectrum of $\delta_d$ (Eq. \ref{eq:fn2d}), 
the 2D power spectrum of the white noise should be at a level of $1$.

The top-left panel of \reffg{fg:ps2d19} shows the results of the data 
without singular mode subtraction.
The power spectrum is peaked at the low-$\tau$ end, which indicates a 
strong correlation across frequency channels. Such strong frequency correlation 
results in a smooth frequency spectrum, which is clearly shown in the waterfall plots
(top panels of \reffg{fig:wfsvd}). The power spectrum has
a tail below $f\sim 10^{-2}$ extending to higher $\tau$. Such a tail structure 
indicates other 1/f components, which are less correlated across frequency channels.
However, the model is only fitted to the component with strong frequency correlation
as it is the dominant one and the model assumes a single spectrum index.
This is also the reason why the temporal power spectrum model can not fit the
data without singular mode subtraction as discussed in \refsc{sec:tpsd}.

The strong frequency correlated component can be removed by subtracting 
the first singular mode. The waterfall plots of the first singular mode
are shown in the top panel of \reffg{fig:wfmod}, which have structures 
consistent with the raw data. The 2D power spectrum of the data with 
first mode subtracted is shown in the top-right panel of \reffg{fg:ps2d19}.
Comparing to the top-left panel, the strong frequency correlated component
is subtracted out. 

With more singular modes subtraction, the 1/f-type power spectrum is 
highly reduced and the knee frequency is also reduced to lower values.
The 2D power spectrum model, \refeq{eq:fn2d}, is fitted to the measurements
for each antenna and polarization with a differing number of SVD modes subtracted,
individually. 
The fitting parameters $f_{\rm k, 20MHz}$, $\alpha$ and
$\beta$ are shown in \reffg{fg:fkab}, where each square marker shows 
the fitting value for one antenna. The results for the two polarizations 
are shown in the left and right panels and the results with different 
singular mode subtraction are shown in different colors as labeled in 
the legend. The median value of the fitting parameters, as well as the r.m.s,
over all dishes are listed in \reftb{tb:fitting}.

As shown in \reffg{fg:fkab}, without singular mode subtraction, the
fitting value for $\beta$ is less than $0.2$, which indicates a 
strong correlation across frequency. After one singular mode is subtracted,
the $\beta$ values increase significantly, but do not keep increasing
as further modes are subtracted. This indicates that the singular modes 
are sensitive to highly correlated structures either in frequency
or time axis. The fitting value for $f_{\rm k, 20MHz}$ is continuously 
reduced with additional singular mode subtraction. 
With one mode subtraction, the fitting values, especially 
$f_{\rm k, 20MHz}$ and $\alpha$, have large scatter between different 
antennas. The scatter indicates that the correlation features are too 
complex to be described by one singular mode and the residuals, which can
be different between antennas, are dominating the second singular mode.
With $2$ modes subtraction, 
the knee frequency at $20\,$MHz is reduced to around $3\times 10^{-3}\,{\rm Hz}$, 
which indicates that the system 1/f-type variations are well under the 
thermal noise fluctuation over $\sim 3\times 10^{2}\,{\rm s}$. The time scale can be
even longer with $5$ modes subtraction. However, as more modes are removed, there is the danger of over-cleaning, although $5$ or more modes removal is standard in foreground cleaning methods. We plan to investigate this behaviour further in follow up work with simulations.

\begin{table*}
    {\scriptsize
    \centering
    \caption{
        The median value of fitting parameters across all dishes. 
        The errors are the r.m.s. of the fitting values across all dishes.
    }\label{tb:fitting}
    \begin{tabular}{c|cc|cc|cc}\hline
                    & \multicolumn{2}{c|}{$\lg f_{\rm k, 20MHz}$}
                    & \multicolumn{2}{c|}{$\alpha$}
                    & \multicolumn{2}{c}{$\beta$} \\ 
                    & HH              & VV              & HH              & VV              & HH              & VV \\ \hline \hline
                    \multicolumn{7}{c}{2D power spectrum density}\\\hline
        1 mode subtraction &$-1.77\pm 0.49$ & $-1.97\pm 0.38$ & $ 0.98\pm 0.25$ & $ 1.00\pm 0.33$ & $ 0.42\pm 0.06$ & $ 0.41\pm 0.06$\\
        2 mode subtraction &$-2.49\pm 0.26$ & $-2.65\pm 0.28$ & $ 1.39\pm 0.25$ & $ 1.39\pm 0.37$ & $ 0.52\pm 0.05$ & $ 0.51\pm 0.08$\\
        5 mode subtraction &$-3.04\pm 0.13$ & $-3.19\pm 0.32$ & $ 1.70\pm 0.35$ & $ 1.80\pm 0.55$ & $ 0.51\pm 0.06$ & $ 0.49\pm 0.11$\\\hline
                    \multicolumn{7}{c}{1D power spectrum density}\\\hline
        1 mode subtraction &$-2.23\pm 0.18$ & $-2.23\pm 0.18$ & $ 2.28\pm 0.34$ & $ 2.25\pm 0.41$ & -- & -- \\
        2 mode subtraction &$-2.40\pm 0.21$ & $-2.40\pm 0.21$ & $ 1.83\pm 0.40$ & $ 1.98\pm 0.38$ & -- & -- \\
        5 mode subtraction &$-2.78\pm 0.14$ & $-2.85\pm 0.21$ & $ 1.28\pm 0.26$ & $ 1.47\pm 0.37$ & -- & -- \\\hline
    \end{tabular}
    }
\end{table*}

The white dotted lines in \reffg{fg:ps2d19wBAO} indicate the corresponding 
cosmological scales projecting to the $f-\tau$ space, assuming observation 
at $900\,{\rm MHz}$ with scan speed of $5\,{\rm arcmin}\,s^{-1}$. The temporal and 
spectroscopic wavenumber, $f$ and $\tau$, are related to the cosmological 
scales $k$ by
\begin{align}
    \tau = \frac{\nu_0}{\nu_{\rm obs}^2} \frac{c}{H(z)} \frac{k_\parallel}{2\pi},\;\;
    f      = \frac{k_\perp\chi(z) u}{2\pi},\;\;{\rm and}\;
    k^2    = k^2_\parallel + k^2_\perp,
\end{align}
in which, $k$ is in units of ${\rm Mpc}^{-1}h$; $\nu_0=1420{\rm MHz}$ 
is the rest frame \hi emission line frequency;
$c$ is the speed of light; $\nu_{\rm obs}$ is the observing frequency 
and $u$ is the scanning speed. We assume the fiducial cosmology parameters
from \citet{2018arXiv180706209P} ($h=0.6736$, $\Omega_{\rm m}=0.3153$ and 
$\Omega_\Lambda = 0.6847$). We see that there is a large region in the $f-\tau$ space that will be available for the 21cm measurements. In particular, even without any mode subtraction, most contamination, either because of 1/f noise or foregrounds is constrained to a region of low $f$ or low $\tau$.

\vspace{-0.1cm}
\section{Summary and Conclusions}\label{sec:conclusion}

In this work we measured the power spectrum density of the 1/f noise
for the MeerKAT receiver system. The analysis is performed with 
South Celestial Pole (SCP) tracking data to avoid
sky variations. Two SCP tracking datasets are used in this analysis.
We find a relatively RFI free frequency range from $1313.6758\,$ MHz 
to $1461.8457\,$ MHz. Absolute flux calibration is ignored in our 
analysis as the data are normalized with the time averaged system temperature
for each frequency channel.

We apply Singular Value Decomposition to the data and determine how effective removing the first several principal components is on suppressing the time-frequency correlated noise. The results show that indeed, the 1/f noise can be drastically reduced by removing the first few SVD modes. Moreover, the correlation features are well described by the proposed
noise model with just a few parameters. Using the parameterisation presented in \refeq{eq:fn2d}, the removal of a single SVD mode reduces $\beta$ from $\sim 0.16$ (e.g., 1/f noise that is highly correlated in frequency) to $\beta = 0.5$, which indicates a large reduction in correlation across frequency. Along the temporal axis of the time-ordered data, a similar reduction is seen, with the raw data averaged over 20 MHz having a knee frequency ($f_k$) of 10$^{-1}$\,Hz, which is reduced to $\sim 3\times 10^{-3}\,{\rm Hz}$ with $2$ mode subtraction, indicating that the system 1/f-type variation is well under the thermal noise fluctuation over a few hundred seconds time scales. The results from this analysis, along with the described noise power spectrum model, can be used in realistic noise simulations for MeerKAT and extended to SKA1-Mid.

The 2d power spectrum shows that the 1/f noise is constrained to a small region of either low $\tau$ or low $f$, e.g. large scale correlations in time or in frequency.
This provides many scales where the 21cm signal can be probed without contamination. With scanning speeds of $5\ {\rm arcmin}\,s^{-1}$, $10^{3}\,{\rm s}$ time scales would correspond to $\sim 80$ degrees, which is enough for our cosmological purposes. Longer time scales can be achieved by using noise diodes. Our calibration plan is to use celestial sources for calibration on timescales $\sim1.5\,{\rm hour}$ (absolute flux and bandpass calibration) and noise diodes for shorter time scales. We can then apply a conservative SVD cleaning on the time-ordered data (e.g. $2$ modes) in order to remove most of the 1/f noise contamination. These modes correspond to large frequency scales where we expect the 21cm correlations to be negligible. Therefore, this cleaning should have a minor impact on the 21cm signal given what we already know about foreground cleaning methods \citep[e.g.][]{2015MNRAS.447..400A}.
Any residual noise will be included in the map making process which will allow for correlated noise in frequency. Finally we will apply foreground cleaning to the maps. 

The 1/f noise has been a substantial challenge to precision radio cosmology in the past and, if it is not carefully treated, has been shown to be detrimental to future HI IM experiments. We have demonstrated here a methodology that can be used to effectively suppress 1/f noise in single dish HI IM observations that should preserve the cosmological signal. In future work we plan demonstrate the effectiveness of this technique both on simulated and real MeerKAT data. 

\section*{Acknowledgments}
We are grateful to Phil Bull, Clive Dickinson and Jonathan Sievers for very useful discussions.
YL and MGS acknowledge support from the South African Square Kilometre Array Project 
and National Research Foundation (Grant No. 84156).
The MeerKAT telescope is operated by the South African Radio Astronomy Observatory, 
which is a facility of the National Research Foundation, an agency of the Department 
of Science and Innovation.
We acknowledge the use of the Inter-University Institute for Data Intensive Astronomy 
(IDIA) and Ilifu computing facilities.

\appendix
\section{ }
\subsection*{Relation between temporal power spectrum density 
and 2D Power spectrum density}

The temporal power spectrum density can be expressed as,
\begin{align}\label{eq:tps1}
S_{ij}(f) = \langle \delta (f, \nu_i) \delta^\dagger (f, \nu_j) \rangle
= \int \dd \nu\, \dd \nu'\, \phi_i(\nu)\phi_j^\dagger(\nu') \xi(f, \nu_{ij})
\end{align}
in which, $\nu_{ij}=\nu_i - \nu_j$, $\phi_i(\nu)$ is the spectroscopic
window function and $\delta(f, \nu_i) = \int \dd \nu\, \phi_i(\nu) \delta(f, \nu)$;
$\xi(f, \nu_{ij})$ is related to the 2D power spectrum density, $S(f, \tau)$, via
inverse Fourier transform,
\begin{align}\label{eq:xi}
    \xi(f, \nu_{ij}) = \int \dd \tau S(f, \tau) \exp\left[2\pi i \tau\nu_{ij}\right].
\end{align}
Substituting \refeq{eq:xi} to \refeq{eq:tps1}, the temporal power spectrum density
can be further expressed as
\begin{align}
S_{ij}(f) = \int \dd \tau\, \phi_i(\tau) \phi^\dagger_j(\tau) S(f, \tau)
\end{align}
where $\phi_i(\tau) = \int \dd \nu\, \phi_i(\nu) \exp[-2\pi i \tau \nu]$
is the Fourier transform of the spectroscopic window function.
If we ignore the cross correlation between frequencies, the diagonal term
of the temporal power spectrum density is,
\begin{align}\label{eq:2d1d}
S_i(f) = \int \dd \tau\, \phi_i^2(\tau) S(f, \tau)
\end{align}

\subsection*{The white noise level}
If we use a top-hat window function with width of $\delta \nu$,
the Fourier transform of the top-hat window function, $\phi_i(\tau)$,
can be expressed with a \sinc~function,
\begin{align}
    \phi_i(\tau) = \sinc(\pi \delta \nu \tau),
\end{align}
where the window function is normalized with
$\int\dd\tau\phi_i^2(\tau) = \int\dd\tau\phi_i(\tau) = 1/\delta\nu$.
Substituting the 2D white noise power spectrum model 
(the first term of \refeq{eq:fn2d}), \refeq{eq:2d1d} becomes,
\begin{align}
    A \int \dd \tau\, \phi_i^2(\tau) = \frac{A}{\delta \nu},
\end{align}
which is consistent with the white noise term of 1D power spectrum density model
\refeq{eq:tmodel}. 

\subsection*{The knee frequency conversion between frequency resolutions}
We firstly model the 2D 1/f noise power spectrum density with $f_0$ at arbitrary
frequency resolution,
\begin{align}
S(f, \tau) = A\left(\frac{f_0}{f}\right)^\alpha
\left(\frac{\tau_0}{\tau}\right)^{\frac{1-\beta}{\beta}}.
\end{align}
Substituting the noise model into \refeq{eq:2d1d},
\begin{align}
    A \left(\frac{f_0}{f}\right)^\alpha \int \dd \tau \phi_i^2(\tau)
    \left(\frac{\tau_0}{\tau}\right)^{\frac{1-\beta}{\beta}}
    = A K \left(\frac{f_0}{f}\right)^\alpha
\end{align}
where $K$ is
\begin{align}
    K = \int \dd \tau \, \sinc^2(\pi \delta \nu \tau)
    \left( \frac{\tau_0}{\tau} \right)^{\frac{1 - \beta}{\beta}}.
\end{align}
The 1D power spectrum model assumes the 1/f noise is
reduced with factor of $\delta \nu$, which indicates that the 1/f noise is
uncorrelated across frequencies. That equivalent to set $\beta=1$ in
the 2D power spectrum density model,
\begin{align}
    AK\left(\frac{f_0}{f}\right)^\alpha &=
    A\left(\frac{f_0}{f}\right)^\alpha\int \dd \tau\,
    \sinc^2(\pi\delta \nu \tau) \left( \frac{\tau_0}{\tau} \right)^{0}
    = \frac{A}{\delta \nu}\left(\frac{f_0}{f}\right)^\alpha.
\end{align}
Comparing with the second term of 1D power spectrum density model \refeq{eq:tmodel},
we have $f_0=f_k$. This indicates that, if the 1/f noise is fully uncorrelated
across frequencies, the knee frequency is constant with any frequency resolution.

On the other hand, if $\beta\to0$, the spectroscopic spectrum density index is
approaching to $+\infty$ and the spectroscopic power spectrum density model becomes
a Dirac delta function, $H(\tau)\to\delta^{\rm D}(\tau)$. In this case, we have
\begin{align}\label{eq:beta0integral}
    AK\left(\frac{f_0}{f}\right)^\alpha &=
    A\left(\frac{f_0}{f}\right)^\alpha\int \dd \tau\,
    \sinc^2(\pi\delta \nu \tau) \delta^{\rm D}(\tau)
    = \frac{A}{\Delta \nu}\left(\frac{f_0}{f}\right)^\alpha.
\end{align}
where $\Delta \nu=1/\tau$. When $\tau=0$, $\Delta \nu\to+\infty$, 
indicating the infinity frequency bandwidth. 

However, the measurements are always 
limited within finite frequency bandwidth. If we write the integrals in
terms of discrete sums, \refeq{eq:2d1d} is expressed as,
\begin{align}
    S_i(f) = \sum_{p=0}^{N_\nu-1} \phi_i^2(p \tau_0) S(f, p\tau_0) \tau_0,
\end{align}
where $\tau_0$ is the minimal spectroscopic frequency interval
and related to the minimal frequency interval $\nu_0$ via
$\tau_0=1/(N_\nu\nu_0)=1/\Delta\nu$, where $\Delta\nu$ is the full frequency
bandwidth. The \refeq{eq:beta0integral} becomes,
\begin{align}
    A\left(\sum_{p=0}^{N_\nu-1} \sinc^2(\pi\delta\nu p\tau_0) \tau_0
    \delta^{D}_p \right) \times
    \left(\frac{f_0}{f}\right)^\alpha
    = \frac{A}{\Delta\nu}\left(\frac{f_0}{f}\right)^\alpha
\end{align}
Comparing with the 1D power spectrum density model, we have,
\begin{align}
    \frac{1}{\Delta \nu}\left(\frac{f_0}{f}\right)^\alpha =
    \frac{1}{\delta \nu}\left(\frac{f_k}{f}\right)^\alpha, \\
    \lg f_0 = \lg f_k - \frac{1}{\alpha} \lg {\frac{\delta \nu}{\Delta \nu}}.
\end{align}
When $\delta \nu = \Delta \nu$, we have $f_0 = f_k$, which indicates that
$f_0$ is the knee frequency at full frequency bandwidth, which is corresponding 
to the minimal spectroscopic frequency interval $\tau_0$.

In the case of $0<\beta<1$, the shift of $f_0$ with frequency resolution is
dependents on the frequency correlation properties.
The relation of $f_k$ between different frequency resolutions, 
$\delta \nu$ and $\delta \nu'$, is expressed as
\begin{align}
    \lg f_{k, \delta \nu} = \lg f_{k, \delta \nu'} + \frac{1}{\alpha}
    \lg \left( \frac{K \delta \nu}{K'\delta \nu'} \right).
\end{align}
Replacing $f_0$ with $f_k$, we have \refeq{eq:fn2d},
\begin{align}
    S(f, \tau) = A \left( 1 + \frac{1}{K\delta\nu}\left(\frac{f_k}{f}\right)^\alpha 
    \left(\frac{\tau_0}{\tau}\right)^{\frac{1-\beta}{\beta}}\right).
\end{align}

\bibliographystyle{mn2e}
\bibliography{main}
\end{document}